\documentclass[aps, prd, 10pt, twocolumn, superscriptaddress,noshowpacs, preprintnumbers, longbibliography,nofootinbib,bibnotes,floatfix]{revtex4-2}
\pdfoutput=1

\usepackage{array}
\usepackage{amsmath}
\usepackage{amsfonts}
\usepackage{amssymb}
\usepackage{mathtools}
\usepackage{graphics}
\usepackage{tabularx}
\usepackage[export]{adjustbox}
\usepackage{multirow}
\usepackage{citesort}
\usepackage{graphicx}
\usepackage{url}
\usepackage{soul}
\usepackage{physics}
\usepackage{bm}
\usepackage[dvipsnames]{xcolor}
\usepackage[utf8]{inputenc}
\usepackage[normalem]{ulem}

\usepackage{tikz,xcolor}
\usepackage{hyperref}

\definecolor{lime}{HTML}{A6CE39}
\DeclareRobustCommand{\orcidicon}{\hspace{-1mm}
 \begin{tikzpicture}
 \draw[lime, fill=lime] (0,0) 
 circle [radius=0.16] 
 node[white] {{\fontfamily{qag}\selectfont \tiny \,ID}};
 \draw[white, fill=white] (-0.0525,0.095) 
 circle [radius=0.007];
 \end{tikzpicture}
 \hspace{-3mm}
}

\foreach \x in {A, ..., Z}{\expandafter\xdef\csname orcid\x\endcsname{\noexpand\href{https://orcid.org/\csname orcidauthor\x\endcsname}
 {\noexpand\orcidicon}}
}

\begin{document}

\title{Electron-neutrino lepton number crossings: \\Variations with the supernova core physics}

\author{Marie Cornelius\orcidA{}}
%\email{marie.cornelius@nbi.ku.dk}
\affiliation{Niels Bohr International Academy and DARK, Niels Bohr Institute, University of Copenhagen, Blegdamsvej~17, 2100 Copenhagen, Denmark}
\author{Irene Tamborra\orcidB{}}
\affiliation{Niels Bohr International Academy and DARK, Niels Bohr Institute, University of Copenhagen, Blegdamsvej~17, 2100 Copenhagen, Denmark}
\author{Malte Heinlein\orcidC{}}
\affiliation{Max-Planck-Institut f\"{u}r Astrophysik, Karl-Schwarzschild-Str.~1, 85748 Garching, Germany}
\affiliation{Technische Universit\"{a}t M\"{u}nchen, TUM School of Natural Sciences, Physics Department, James-Franck-Str.~1, 85748 Garching, Germany}
\author{Shashank Shalgar\orcidD{}}
\affiliation{Niels Bohr International Academy and DARK, Niels Bohr Institute, University of Copenhagen, Blegdamsvej~17, 2100 Copenhagen, Denmark}
\author{Hans-Thomas Janka\orcidE{}}
\affiliation{Max-Planck-Institut f\"{u}r Astrophysik, Karl-Schwarzschild-Str.~1, 85748 Garching, Germany}

\date{\today}

\begin{abstract}
A crucial ingredient affecting fast neutrino flavor conversion in core-collapse supernovae (SNe) is the shape of the angular distribution of the electron-neutrino lepton number (ELN). The presence of an ELN crossing signals favorable conditions for flavor conversion. However, the dependence of ELN crossings on the SN properties is only partially understood. We investigate a suite of 12 spherically symmetric neutrino-hydrodynamics simulations of the core collapse of a SN with a mass of $18.6 M_\odot$; each model employs different microphysics (i.e., three different nuclear equations of state, with and without muon creation) and includes or not a mixing-length treatment for proto-neutron star convection. We solve the Boltzmann equations to compute the neutrino angular distributions relying on static fluid properties extracted from each of the SN simulations in our suite for six selected post-bounce times. We explore the dependence of the ELN distributions on the SN microphysics and proto-neutron star convection. We find that the latter shifts the proto-neutron star radius outwards, favoring the appearance of ELN crossings at larger radii. On the other hand, muon creation causes proto-neutron star contraction, facilitating the occurrence of ELN crossings at smaller radii. These effects mildly depend on the nuclear equation of state. Our findings highlight the subtle impact of the SN microphysics, proto-neutron star convection, and neutrino transport on the ELN angular distributions.
\end{abstract}

\maketitle

\section{Introduction}
Core-collapse supernovae (SNe) are among the most energetic transient events that occur in our Universe. Neutrinos are abundantly produced during the core collapse of massive stars and play a key role in the transport of energy and lepton number, as well as in driving the explosion according to the delayed neutrino-driven mechanism, cf.~Refs.~\cite{Burrows:2020qrp,Tamborra:2024fcd,Vitagliano:2019yzm} for recent reviews. Multi-dimensional hydrodynamical SN simulations have reached a mature stage, with a wide range of progenitor properties being explored~\cite{Janka:2025tvf,Burrows:2016ohd}. However, due to its complexity, neutrino transport still involves a number of approximations. First, in most state-of-the-art simulations of stellar collapses, the evolution of the neutrino field is tracked relying on the lowest angular moments of the neutrino distribution functions, with closure schemes being employed to minimize the error induced by neglecting the higher angular moments (cf., e.g., Ref.~\cite{Mezzacappa:2020oyq} and references therein). Second, neutrino flavor conversion is not taken into account in state-of-the-art SN simulations. However, a growing body of work points out that flavor evolution due to neutrino-neutrino refraction should already occur in the proximity of neutrino decoupling~\cite{Tamborra:2020cul,Volpe:2023met,Johns:2025mlm}. 
Recent work, adopting simplified approaches to gauge the impact of flavor conversion in hydrodynamic simulations, points out that the change of flavor of neutrinos can affect the SN explosion mechanism, the multi-messenger observables, and nucleosynthesis~\cite{Nagakura:2023mhr,Ehring:2023abs,Ehring:2023lcd,Mori:2025cke,Wang:2025nii}.

At high neutrino densities, fast flavor conversion can take place~\cite{Sawyer:2005jk,Sawyer:2015dsa,Chakraborty:2016yeg,Chakraborty:2016lct}, if a crossing occurs between the angular distributions of electron neutrinos and antineutrinos--the electron lepton number (ELN) crossing--under the assumption that the non-electron flavors have identical properties~\cite{Izaguirre:2016gsx,Morinaga:2021vmc,Padilla-Gay:2021haz,2023PhRvD.107l3024F}. Deep in the SN core, where the electron flavors are in chemical equilibrium with the stellar medium, crossings in the ELN angular distributions may develop in the proximity of the rapid drop of the electron fraction (deleptonization). The latter is responsible for a shift of beta equilibrium towards negative chemical potentials for $\nu_e$'s. Due to differences in the interaction rates of electron neutrinos and antineutrinos with the stellar medium, $\bar\nu_e$'s decouple from the matter background at smaller radii than $\nu_e$'s, exhibiting more forward-peaked distributions~\cite{Tamborra:2017ubu,Brandt:2010xa,Shalgar:2019kzy}.

The neutrino angular distributions have been computed employing Boltzmann neutrino transport for a small set of SN hydrodynamic simulations, cf.~e.g.,~Refs.~\cite{Sumiyoshi:2012za,Akaho:2020xgb,Nagakura:2017mnp,Brandt:2010xa,Fischer:2009af,Tamborra:2017ubu}. Therefore, in most cases, the neutrino angular distributions are reconstructed employing moment-based schemes to assess the existence of ELN crossings in SN simulations, cf.~e.g.,~Refs.~\cite{Abbar:2020fcl,Abbar:2020qpi,Capozzi:2020syn,Dasgupta:2018ulw,Glas:2019ijo,Nagakura:2021hyb}.
Although Refs.~\cite{Johns:2021taz,Cornelius:2025tyt} have recently questioned the inference of the ELN properties from SN simulations based on moment-based schemes, ELN crossings appear to be a widespread phenomenon in SNe with characteristic features depending on the radial regions where they develop~\cite{Nagakura:2021hyb}. These findings suggest that ELN crossings may depend on the SN dynamics as well as on the SN microphysics. 

This paper explores the conditions under which ELN crossings occur in a suite of spherically symmetric SN models. Our goal is to assess how proto-neutron star (PNS) convection and SN microphysics [e.g.,~the presence of muons and the nuclear equation of state (EoS)] influence the formation and characteristics of these crossings. To this end, we adopt $12$ spherically symmetric core-collapse simulations of an $18.6 M_{\odot}$ progenitor, which were performed with and without muons and with and without a mixing-length treatment of PNS convection, and three different nuclear EoSs. For selected post-bounce time snapshots, following the approach presented in Refs.~\cite{Shalgar:2023aca,Shalgar:2024gjt,Shalgar:2025oht}, we extract static radial profiles of the main fluid properties from the SN simulations (such as the baryon density, medium temperature, electron and neutron fractions, and chemical potentials) to solve the Boltzmann equations in the surroundings of the neutrino decoupling region and compute the neutrino angular distributions. 

Our work is organized as follows. We introduce the SN models employed in our analysis in Sec.~\ref{Sec:SN_simulations}. Section~\ref{Sec:Boltzmann} outlines the approach adopted to compute the angular distributions for all flavors by solving the Boltzmann equations. We also explore the main features and radial evolution of the neutrino angular distributions. A comparison between the features of the ELN angular distributions and the appearance of ELN crossings across our SN models is presented in Sec.~\ref{sec:comparison}. We then discuss our findings and provide an outlook in Sec.~\ref{sec:conclusions}. In addition, in Appendix.~\ref{appendix:comparison}, we compare the first two angular moments extracted from the SN simulations with those computed by solving our Boltzmann equations; we also investigate the impact of the energy resolution on the neutrino angular distributions. 

\section{Core-collapse supernova models}
\label{Sec:SN_simulations}
We adopt a suite of spherically symmetric SN models with a progenitor mass of $18.6 M_{\odot}$~\cite{Garching_CCSN_archive} and PNS baryonic mass of $1.6 M_\odot$. These models employ three different nuclear EoSs: 
the Lattimer and Swesty EoS with incompressibility modulus equal to $220$~MeV (LS220)~\cite{Lattimer:1991nc}, the Steiner, Fischer, and Hempel (SFHo) EoS~\cite{Steiner:2012rk,Hempel:2009mc}, and the Typel et al.~(DD2) EoS~\cite{Typel:2009sy,Hempel:2009mc,Hempel:2011mk}. The selected EoSs encompass a range of nuclear-matter and PNS properties, with the LS220 and SFHo EoSs generally considered to be soft EoSs and the DD2 EoS being stiffer. 

Due to the relevance of muons for the PNS structure and in the explosion mechanism~\cite{Bollig:2017lki}, muon production is taken into account in half of our SN models following Refs.~\cite{Bollig:2017lki,Capozzi:2020syn}. Muon creation effectively softens the EoS, triggering faster PNS contraction and affecting neutrino decoupling. For the remaining half of the SN models, muons are not included and the emission properties of the muon and tau flavors are indistinguishable ($\nu_x=\nu_\mu$ or $\nu_\tau$, and $\bar\nu_x=\bar\nu_\mu$ or $\bar\nu_\tau$), except for small differences between the heavy-lepton neutrino and antineutrinos due to weak-magnetism corrections in the neutral current scatterings on nucleons.
Proto-neutron star convection plays a crucial role in the evolution of the electron lepton number and energy loss, accelerating the cooling and the deleptonization of the PNS~\cite{2012PhRvL.108f1103R,Pascal:2022qeg,Mirizzi:2015eza}. To investigate whether PNS convection may affect the development of ELN crossings, half of our models feature a mixing-length treatment for convection~\cite{2012PhRvL.108f1103R,Mirizzi:2015eza}.

These variations allow us to isolate the effects of the microphysics and hydrodynamics on the formation of ELN crossings. A summary of the SN models considered in this paper is provided in Table~\ref{Tab:all_models}.
Hereafter, we adopt the SN model with the LS220 EoS without muons and without PNS convection ($1.61$-LS220) as our benchmark model. Note that the models $1.61$-LS220, $1.61$-LS220$+$c, and $1.62$-LS220$+$c$+$m correspond to Models 1, 2, and 3 of Ref.~\cite{Cornelius:2025tyt}, respectively.

\begin{table}[]
\caption{Overview of the properties of the $12$ SN models adopted in this paper. Columns from left to right indicate the model name, the EoS, whether the SN model takes into account muon production, and whether a mixing-length treatment for PNS convection is considered.}
\renewcommand{\arraystretch}{1.2}
\setlength{\tabcolsep}{5pt}
\begin{tabular}{|l|l|l|l|}
\hline
 SN model & EoS & Muons & PNS conv. \\ \hline \hline
$1.61$-LS220 (benchmark) & LS220 & No & No \\
\hline
$1.61$-LS220$+$c & LS220 & No & Yes \\ \hline
$1.61$-LS220$+$m & LS220 & Yes & No \\\hline
$1.62$-LS220$+$c$+$m & LS220 & Yes & Yes \\\hline
$1.62$-SFHo & SFHo & No & No \\ \hline
$1.62$-SFHo$+$c & SFHo & No & Yes \\ \hline
$1.62$-SFHo$+$m & SFHo & Yes & No \\ \hline
$1.62$-SFHo$+$c$+$m & SFHo & Yes & Yes \\ \hline
$1.62$-DD2 & DD2 & No & No \\ \hline
$1.62$-DD2$+$c & DD2 & No & Yes \\ \hline
$1.62$-DD2$+$m & DD2 & Yes & No \\ \hline
$1.62$-DD2$+$c$+$m & DD2 & Yes & Yes \\ \hline
\end{tabular}
\label{Tab:all_models}
\end{table}

The six-species (four-species for the models without muons) neutrino transport module of the {\tt PROMETHEUS-VERTEX} neutrino-hydrodynamics code used in the SN calculations includes a Boltzmann transport module that also provides the angle-dependent neutrino phase-space distributions (see Ref.~\cite{Tamborra:2017ubu}). However, these are not routinely evaluated (because of the large amount of time-dependent data) and have not been stored for the suite of models used here.
The {\tt VERTEX} neutrino transport solver integrates the velocity-dependent energy and momentum equations for neutrinos and antineutrinos, discretizing them in neutrino energy, space, and time. The angular-moment equations are closed using a variable Eddington factor obtained from the Boltzmann transport module~\cite{Rampp:2002bq}. We refer the interested reader to Sec.~II of Ref.~\cite{Fiorillo:2023frv} for additional details on the SN simulation suite.

\begin{figure*}[t]
    \includegraphics[width=0.99\textwidth]{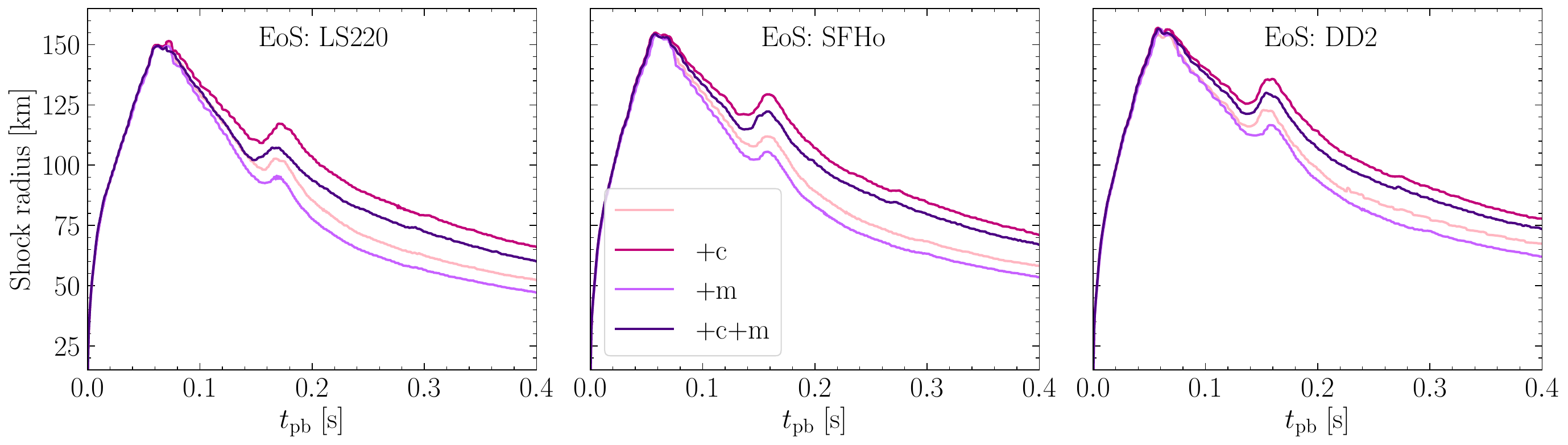}
    \caption{Temporal evolution of the shock radius for our $12$ SN models. Each panel represents the evolution of the shock radius for the SN models employing the same EoS: LS220, SFHo, and DD2, from left to right, respectively.}
    \label{Fig:shock}
\end{figure*}
Figure~\ref{Fig:shock} shows the temporal evolution of the shock radius before the explosion is artificially triggered. The SN models are grouped by EoS: LS220, SFHo, and DD2, from left to right, respectively. Independent of the EoS, we see that the shock radius is larger in the models featuring PNS convection and is smaller in the models including muons as a consequence of the PNS contraction. The explosions were triggered (by artificially reducing the matter density in the infall region ahead of the stalled shock) at a time when the PNS reached the desired baryonic mass of $\sim 1.62 M_\odot$. Depending on the SN model, outward shock expansion thus sets in between $430$ and $480$~ms after core bounce.

\begin{figure*}[t]
    \includegraphics[width=0.99\textwidth]{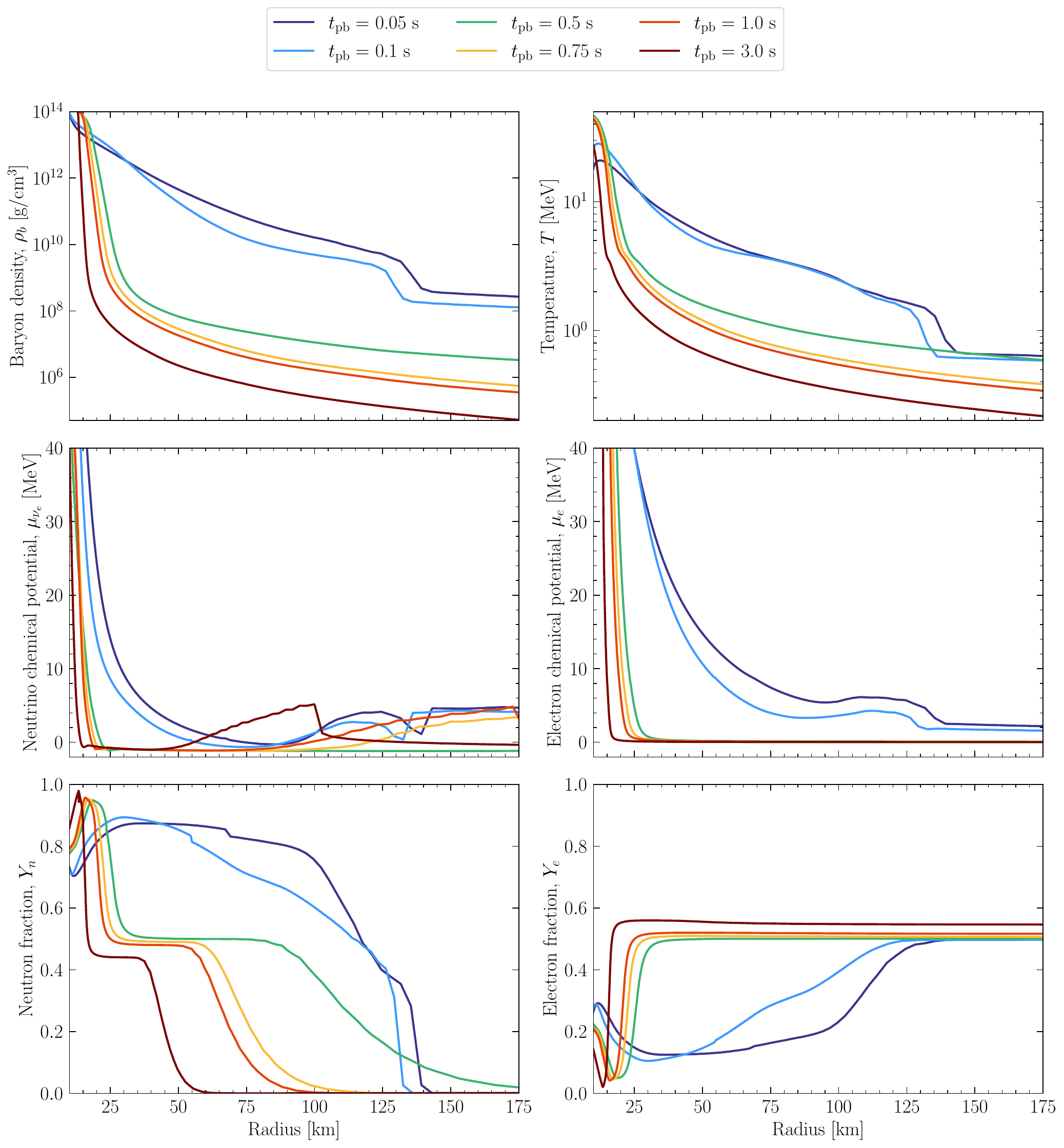}
    \caption{Radial profiles (only shown for radii between $10$ and $175$~km, corresponding to the radial range of our Boltzmann solution; cf.~Table~\ref{Tab:r_ranges}) of the characteristic fluid properties extracted from our benchmark SN model ($1.61$-LS220) at the post-bounce times $t_{\mathrm{pb}}=0.05$, $0.1$, $0.5$, $0.75$, $1$, and $3$~s and adopted to compute the neutrino angular distributions. From top left to bottom right, the different panels represent the baryon number density, the plasma temperature, the electron neutrino and electron chemical potentials, and the neutron and electron fractions.}
    \label{Fig:hydro}
\end{figure*}

In the following, in order to investigate the temporal evolution of the neutrino angular distributions, we select six post-bounce times for all SN simulations: $0.05$, $0.1$, $0.5$, $0.75$, $1$, and $3$~s. These time snapshots span from the early PNS accretion phase through to the Kelvin-Helmholtz cooling phase. The radial profiles of the characteristic fluid properties extracted for our benchmark SN model ($1.61$-LS220) at the selected post-bounce times are displayed in Fig.~\ref{Fig:hydro}.

\section{Neutrino angular distributions}
\label{Sec:Boltzmann}
As mentioned above, the angular distributions of neutrinos and antineutrinos computed by means of the neutrino-hydrodynamic simulations with {\tt PROMETHEUS-VERTEX} were not evaluated and stored for the considered set of SN simulations. References~\cite{Johns:2021taz,Cornelius:2025tyt} highlight the importance of solving the neutrino Boltzmann equations in order to determine the occurrence of ELN crossings, rather than relying on approximate schemes based on the first few angular moments. In this section, we present how we compute the neutrino angular distributions by solving the Boltzmann equations in the decoupling region. We do so by adopting static fluid properties extracted for selected post-bounce times from our SN models.

\subsection{Boltzmann neutrino transport}

The neutrino and antineutrino radiation fields are modeled using the density matrices $\rho(r,\cos\theta,E, t)$ and $\bar\rho(r,\cos\theta,E, t)$; here, $r$, $\cos\theta$, $E$, and $t$ denote the radial direction, the propagation angle relative to $r$, the neutrino energy, and time, respectively. The diagonal terms of the density matrices represent the occupation numbers of neutrinos of different flavors, with the local number density of the flavor $\nu_\alpha$ being defined as $n_{\nu_\alpha}(r) = \int dE d\cos\theta \rho_{\alpha\alpha}(r,\cos\theta,E, t)$. The off-diagonal elements of the density matrices can describe the coherence between flavors, but flavor conversion is not taken into account in this work, and therefore, such terms are zero. Note that, in the solution of the Boltzmann equations, we work in the two-flavor approximation, with $\alpha = e$ or $x$ (with $x$ standing for  $\mu$ or $\tau$ flavors). The Boltzmann equations describing the evolution of the neutrino and antineutrino fields are given by~\cite{Sigl:1993ctk}
\begin{equation}
    \begin{split}
    \left(\pdv{t}+\vec c \cdot \vec \nabla\right) \rho(r,\cos\theta,E,t) = \mathcal{C}[\rho(r,\cos\theta,E,t)]\ , \\
    \left(\pdv{t}+\vec c \cdot \vec \nabla\right) \bar\rho(r,\cos\theta,E,t) = \bar{\mathcal{C}}[\bar\rho(r,\cos\theta,E,t)]\ ,
    \end{split}
\label{Eq:KE}
\end{equation}
where $\vec{c}$ denotes the velocity vector of the neutrinos.

In spherical symmetry, the neutrino propagation term on the left-hand side of Eqs.~(\ref{Eq:KE}) is given by
\begin{equation}
    \Vec{c}\cdot \vec\nabla = 
    \frac{\partial}{\partial \cos{\theta}}\frac{\sin^2\theta}{r} + \cos{\theta}\frac{\partial}{\partial r}\ .
\label{Eq:adv}
\end{equation}
The collision operators on the right-hand side of Eqs.~(\ref{Eq:KE}) take into account (anti)neutrino interactions with matter (i.e., beta processes, pairwise annihilation and production by electron-positron annihilation, nucleon-nucleon bremsstrahlung, and direction-changing interactions). 
Our modeling of the collisional kernel is different from the one in {\tt PROMETHEUS-VERTEX} and is heavily inspired by Ref.~\cite{OConnor:2014sgn} (we refer the interested reader to Sec.~II and Appendix A of~\cite{Shalgar:2023aca}, Sec.~2 of Ref.~\cite{Shalgar:2024gjt}, and references therein for additional details).
The collision term of Eqs.~(\ref{Eq:KE}) does not include muons; hence, we consider the average of the muon and tau flavors.

To compute the neutrino angular distributions for selected post-bounce times, we numerically solve Eqs.~(\ref{Eq:KE}), employing as input the static profiles extracted from our suite of SN models and, as an example, showcased in Fig.~\ref{Fig:hydro} for our benchmark SN model.
Equations~(\ref{Eq:KE}) are solved in a spherical shell surrounding the SN core, spanning from $r_{\min}$ through $r_{\max}$. The width of the shell is tailored for each post-bounce time and SN model to make sure that the Boltzmann equations are solved in a radial range covering the decoupling region for all neutrino energies. Hence, the neutrino distributions are isotropic at $r_{\min}$, and neutrinos are fully decoupled from the stellar medium and stream in the forward direction (with negligible backward flux) at $r_{\max}$. Table~\ref{Tab:r_ranges} provides a summary of the radial ranges used for all SN models in this work. To solve the Boltzmann equations, we use $100$ energy bins in $[1,100]$~MeV, $300$ $\cos\theta$ angular bins, and $150$ radial bins in [$r_{\min}, r_{\max}$], all linearly spaced. 

The neutrino distributions are obtained by solving Eqs.~(\ref{Eq:KE}) until a steady state is reached. Since the boundary value problem, obtained by setting the time-derivative term in Eqs.~(\ref{Eq:KE}) to zero, consists of a set of linear equations upon discretization of the derivative terms, we solve the system of equations using the LU decomposition method. One Newton-Raphson iteration is needed; nonetheless we use three iterations to account for the floating-point error that arises due to the LU decomposition of our large system of equations.
The computational cost of this approach scales as the cube of the number of coupled equations. For the phase-space resolution and small spatial domain considered here, the time-independent solution is significantly faster than the time-dependent one solved in Refs.~\cite{Shalgar:2023aca,Shalgar:2024gjt}. This is because we ignore the coupling between the energy bins, which would make the matrix $100$ times larger.

\begin{table}[]
  \caption{Radial ranges employed to solve the Boltzmann equations and compute the neutrino angular distributions, for each of the SN models in our suite and fixed post-bounce time. Note that we select the same $r_{\min}$ and $r_{\max}$ for simulations with and without muons.}
\renewcommand{\arraystretch}{1.2}
\setlength{\tabcolsep}{4pt}
\begin{tabular}{|l|>{\centering\arraybackslash}p{1.5cm}>{\centering\arraybackslash}p{1.5cm}|>{\centering\arraybackslash}p{1.65cm}>{\centering\arraybackslash}p{1.65cm}|}
\hline
 & \multicolumn{2}{c|}{$1.61$-LS220($+$m)} & \multicolumn{2}{c|}{$1.61(1.62)$-LS220$+$c($+$m)} \\ \hline
$t_{\mathrm{pb}}$ & $r_{\min}$ & $r_{\max}$ & $r_{\min}$ & $r_{\max}$ \\ \hline
0.05 s & 25 km & 175 km & 25 km & 175 km \\
0.10 s & 25 km & 135 km & 25 km & 135 km \\
0.50 s & 15 km & 37 km & 18 km & 37 km \\
0.75 s & 13 km & 31 km & 16 km & 31 km \\
1.00 s & 12 km & 29 km & 14 km & 29 km \\
3.00 s & 10 km & 21 km & 12 km & 21 km \\ \hline \hline

 & \multicolumn{2}{c|}{$1.62$-SFHo($+$m)} & \multicolumn{2}{c|}{$1.62$-SFHo$+$c($+$m)} \\ \hline
$t_{\mathrm{pb}}$ & $r_{\min}$ & $r_{\max}$ & $r_{\min}$ & $r_{\max}$ \\ \hline
0.05 s & 25 km & 175 km & 25 km & 175 km \\
0.10 s & 25 km & 135 km & 25 km & 135 km \\
0.50 s & 17 km & 39 km & 20 km & 39 km \\
0.75 s & 15 km & 33 km & 18 km & 33 km \\
1.00 s & 13 km & 30 km & 15 km & 30 km \\
3.00 s & 11 km & 22 km & 11 km & 22 km \\ \hline \hline

 & \multicolumn{2}{c|}{$1.62$-DD2($+$m)} & \multicolumn{2}{c|}{$1.62$-DD2$+$c($+$m)} \\ \hline
$t_{\mathrm{pb}}$ & $r_{\min}$ & $r_{\max}$ & $r_{\min}$ & $r_{\max}$ \\ \hline
0.05 s & 25 km & 175 km & 25 km & 175 km \\
0.10 s & 25 km & 135 km & 25 km & 135 km \\
0.50 s & 19 km & 41 km & 22 km & 41 km \\
0.75 s & 17 km & 35 km & 19 km & 35 km \\
1.00 s & 15 km & 32 km & 17 km & 32 km \\
3.00 s & 12 km & 23 km & 12 km & 23 km \\ \hline
\end{tabular}
\label{Tab:r_ranges}
\end{table}

We stress that the collisional kernel employed to compute the neutrino angular distributions is different from the one employed in the {\tt PROMETHEUS-VERTEX} neutrino-hydrodynamics code (see Refs.~\cite{Rampp:2002bq,Buras:2005rp,Janka:2006fh,Janka:2012wk} and the updates summarized in Ref.~\cite{Bollig:2017lki} and Sec.~II.A of Ref.~\cite{Fiorillo:2023frv} for details on the modeling of the neutrino reactions in {\tt VERTEX}). Although our goal is not to reproduce the output of hydrodynamic simulations with the highest possible accuracy, we have tested that the neutrino number densities for all post-bounce times and radii differ at most by $\mathcal{O}(10)\%$ between our Boltzmann solution and the output of the SN simulations, as illustrated in Appendix~\ref{appendix:comparison}.
Despite the differences between the transport results from the Boltzmann solvers used here and in the {\tt PROMETHEUS-VERTEX} code used for the SN simulations (see Appendix~\ref{appendix:comparison}), we can investigate the influence of different physics in the SN core by comparing the results obtained with the Boltzmann scheme described in the section.

\subsection{Features of the neutrino angular distributions} 
\label{sec:ELN_dis}
\begin{figure}
    \centering
    \includegraphics[width=0.99\linewidth]{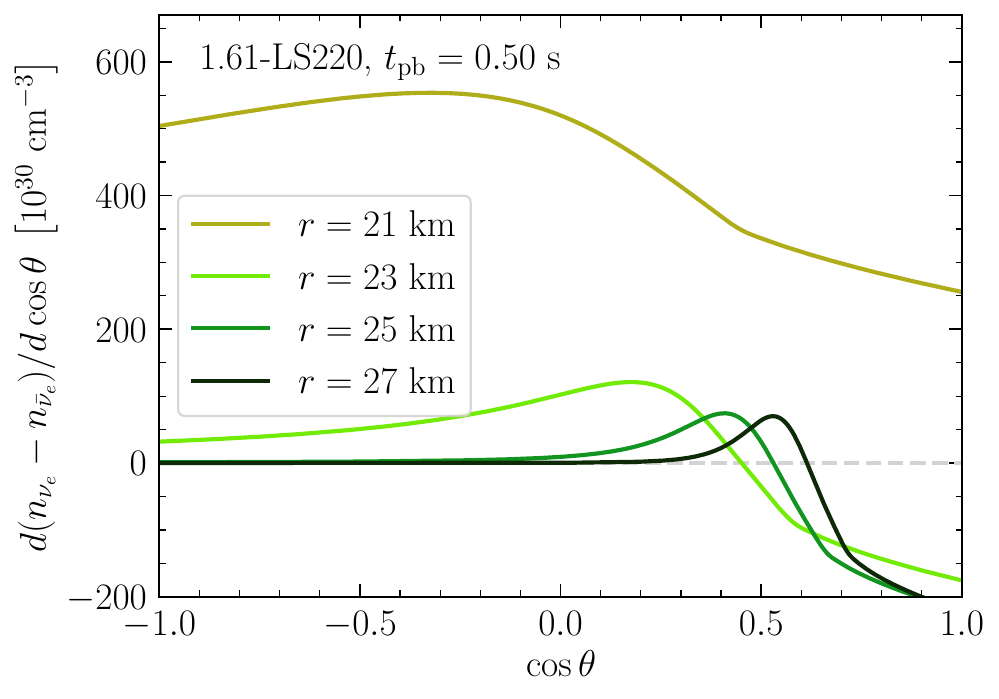}
    \caption{Angular distributions of the energy-integrated differential ELN number density for our benchmark SN model ($1.61$-LS220) at $t_{\mathrm{pb}} = 0.5$~s and for four different radii: $r = 21$~km (olive), $r = 23$~km (light green), $r = 25$~km (medium green), and $r = 27$~km (dark green). To guide the eye, the gray dashed line marks the loci where the ELN crossings should occur. The ELN distribution is isotropic before neutrino decoupling; as the radius increases, the ELN distribution becomes more and more forward peaked and ELN crossings develop.}
    \label{Fig:ELN_radii}
\end{figure}
Figure~\ref{Fig:ELN_radii} shows the ELN angular distributions obtained by solving Eqs.~(\ref{Eq:KE}) for our benchmark SN model at the post-bounce time $t_{\mathrm{pb}}=0.5$~s. To inspect the radial evolution of the ELN angular distribution, we consider four representative radii: $r=21$~km (olive), $r=23$~km (light green), $r=25$~km (green), and $r=27$~km (dark green). We see that the ELN distribution becomes increasingly forward peaked as the radius increases. Moreover, ELN crossings are initially not present and develop as the angular distributions of $\nu_e$'s and $\bar\nu_e$'s become forward peaked, with the location of the ELN crossing moving to larger $\cos\theta$ as the radius increases~\cite{Tamborra:2017ubu,Shalgar:2019kzy}.

\begin{figure*}
    \centering
    \includegraphics[width=0.99\textwidth]{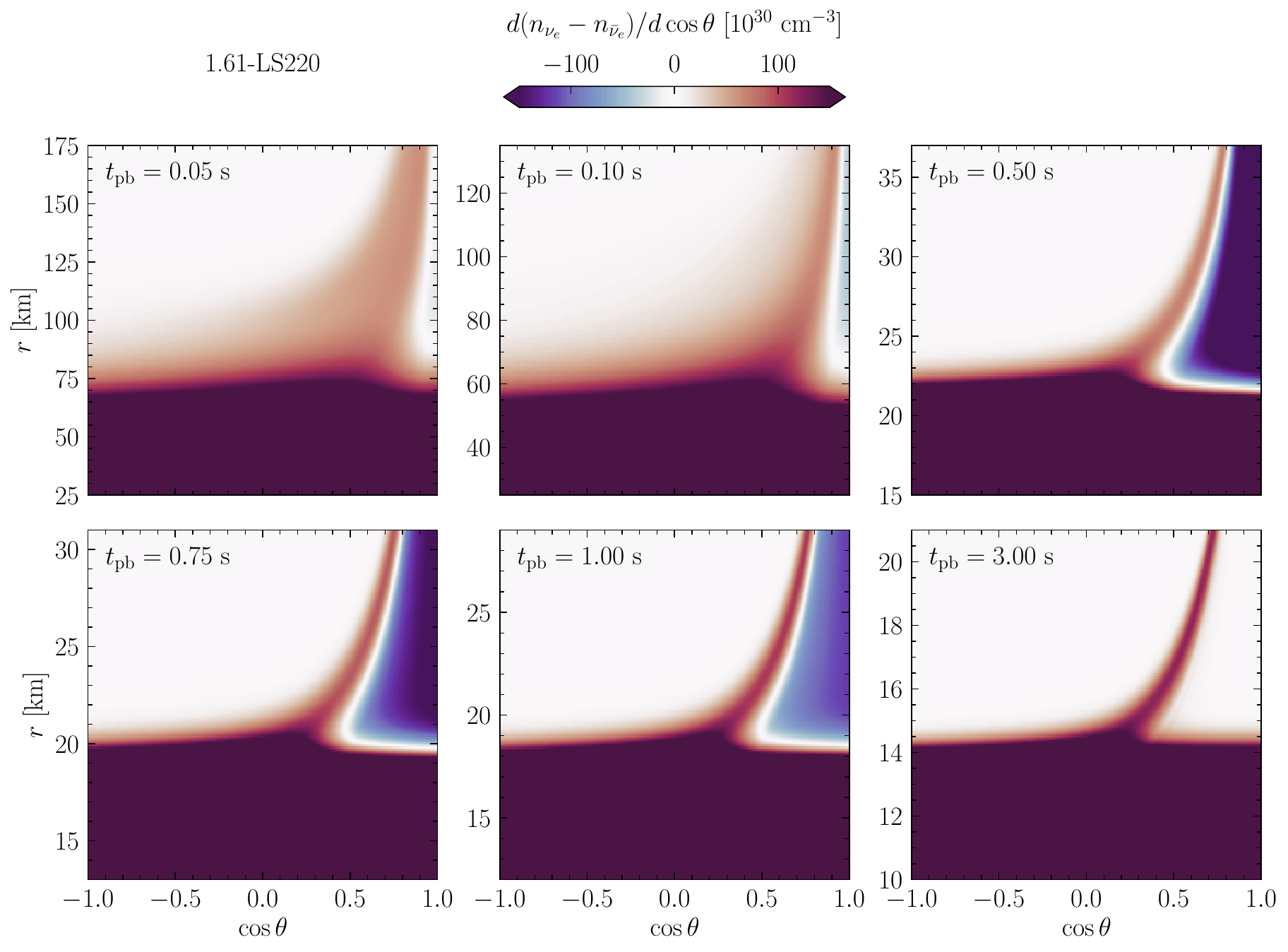}
    \caption{Heatmaps of the energy-integrated ELN number density in the plane spanned by $\cos\theta$ and $r$ for our benchmark SN model ($1.61$-LS220) and for the six selected post-bounce times ($t_{\mathrm{pb}}=0.05$, $0.1$, $0.5$, $0.75$, $1.0$, and $3.0$~s from top left to bottom right, respectively). ELN crossings appear at all post-bounce times, except for $t_{\mathrm{pb}}=3$~s, exhibiting time-dependent features. }
    \label{Fig:contour}
\end{figure*}

Figure~\ref{Fig:contour} shows the evolution of the ELN distribution as a function of the post-bounce time. It displays heatmaps of the energy-integrated ELN number density in the plane spanned by the propagation angle and radius, for our six selected post-bounce times for the benchmark SN model.
The red (blue) regions correspond to an excess of electron neutrinos (antineutrinos). The white contours in the forward direction ($\cos\theta>0$) between the red and blue regions highlight the loci where the ELN changes sign. We can see that the ELN angular distribution appears to be isotropic at small radii and tends to become forward peaked as (anti)neutrino decoupling from matter takes place. As expected, the decoupling radius decreases as $t_{\rm pb}$ increases. Notably, ELN crossings are observed for almost all post-bounce time snapshots, but they exhibit time-dependent features; first (for $t_{\rm{pb}} \lesssim 0.5$~s), as the $\nu_e$ and $\bar\nu_e$ properties largely differ from each other, the ELN crossings tend to be more skewed; then, as $t_{\rm{pb}}$ increases, the ELN crossings become milder as the $\nu_e$ and $\bar\nu_e$ emission properties approach each other, until the ELN crossing disappears at $t_{\rm{pb}} \simeq 3$~s.

\section{Electron-lepton number crossings: comparison between models}
\label{sec:comparison}
\begin{figure*}
    \centering
    \includegraphics[width=0.99\textwidth]{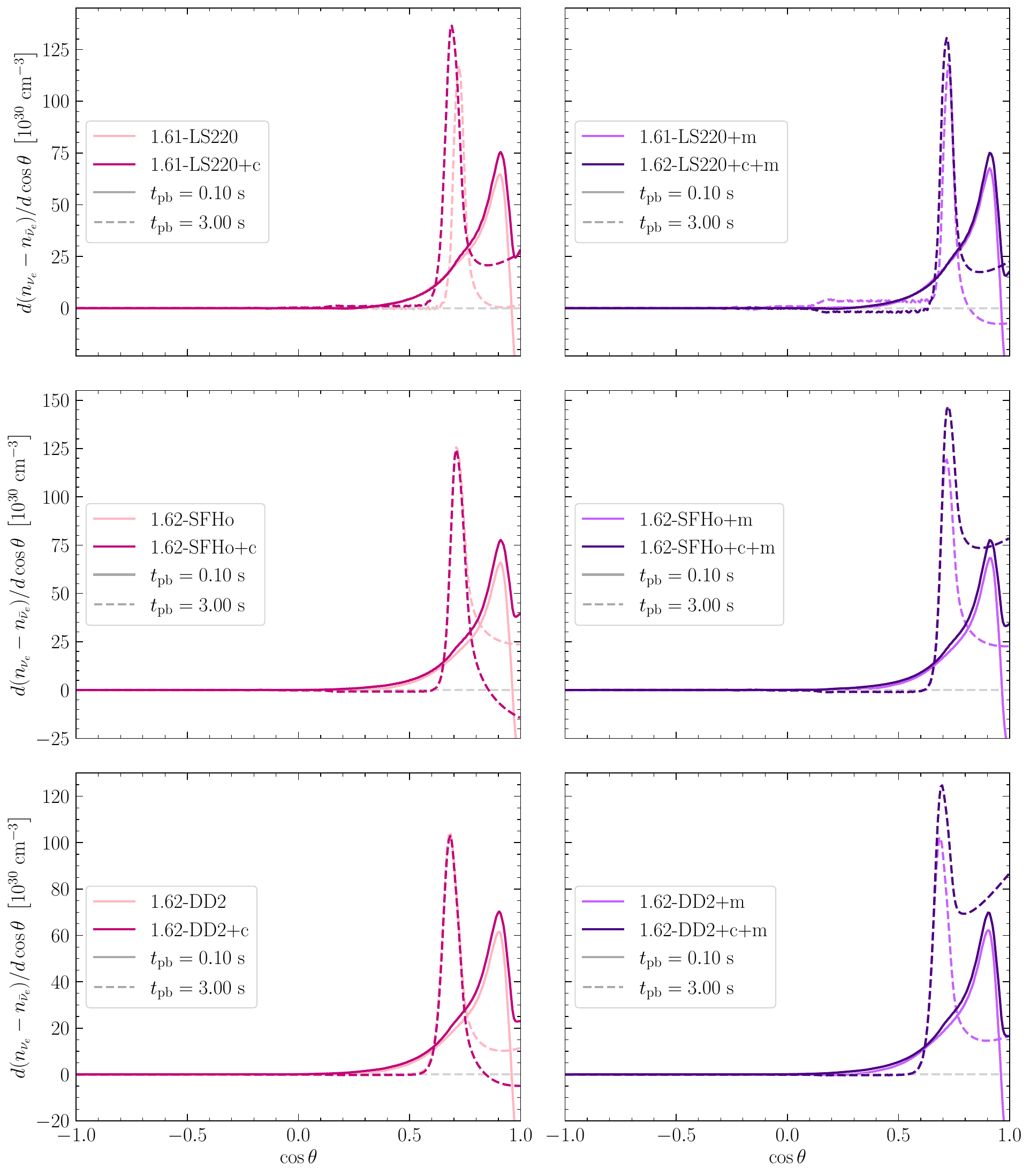}
    \caption{Energy-integrated ELN angular distributions extracted in the proximity of $r_{\max}$ (cf.~main text for details) for the SN models with LS220 (top panels), SFHo (middle panels), and DD2 (bottom panels) EoSs. The left panels show the SN models without muons (and with/without convection), and the right panels display the SN models with muons (and with/without convection).
    For each SN model, the angular distributions are plotted at $t_{\mathrm{pb}}=0.1$~s (solid; $133.2$~km) and $3$~s (dashed; $20.8$~km for the models with EoS LS220, $21.8$~km for the models with SFHo EoS, and $22.8$~km for the models with DD2 EoS). Proto-neutron star convection tends to broaden the shape of the ELN distribution in the forward direction because it leads to larger PNS radii and thus shifts the neutrinospheres to larger radii. Muons negligibly affect the shape of the angular distributions at $0.1$~s, but impact the ones at $3.0$~s. 
    }
    \label{Fig:ELN}
\end{figure*}

The evolution of the ELN angular distribution as a function of radius and post-bounce time, illustrated in Sec.~\ref{sec:ELN_dis} for the benchmark model, is qualitatively similar for all $12$ SN models considered in this work. We now explore how the shape of the ELN distribution and the strength of the ELN crossings vary across our $12$ SN models. To this end, Fig.~\ref{Fig:ELN} shows the energy-integrated ELN distributions for our models, grouped by EoS and extracted at $t_{\mathrm{pb}}=0.1$~s (solid) and $t_{\mathrm{pb}}=3$~s (dashed); we select these two post-bounce times, as the ELN angular distributions display the biggest differences in terms of the existence of crossings. For each model, the angular distributions are extracted in the proximity of $r_{\rm{max}}$, where they are forward peaked (we choose to plot the angular distributions at the radius corresponding to two radial bins before the $r_{\rm{max}}$ one; by doing so, we avoid that the angular distributions may be affected by any numerical fluctuations caused by the upper boundary).

For the LS220 EoS (Fig.~\ref{Fig:ELN}, top panels), ELN crossings in the forward direction appear at $0.1$~s in the proximity of $\mu=1$ for the two SN models without convection (pink and violet lines). The two models with convection (magenta and purple) also feature crossings, but these are less prominent and appear between $\mu=0$ and $\mu=0.5$. At $3.0$~s, the two models with muons (violet, purple) show an ELN crossing.
A similar trend can also be observed for the models with SFHo and DD2 EOSs (middle and bottom panels of Fig.~\ref{Fig:ELN}), although for $t_{\mathrm{pb}}=3.0$~s it is instead only the $+$c model that exhibits a clear crossing. 
In general, independently of the EoS, PNS convection leads to a larger PNS radius and thus larger neutrinosphere radii; as a consequence, in the presence of PNS convection we find a broader shape of the ELN angular distributions. 
The nuclear EoS negligibly affects the depth of the ELN crossings, while it affects the depth of the tail of the ELN distribution near $\mu=1$.
Muons do not impact the ELN distributions at $0.1$~s. However, they play a role in the shape of the ELN distribution at $3.0$~s; here the models with convection and muons (purple) have noticeably different ELN distributions than those with convection and without muons (magenta).

\begin{figure*}[h!]
    \includegraphics[width=0.99\textwidth]{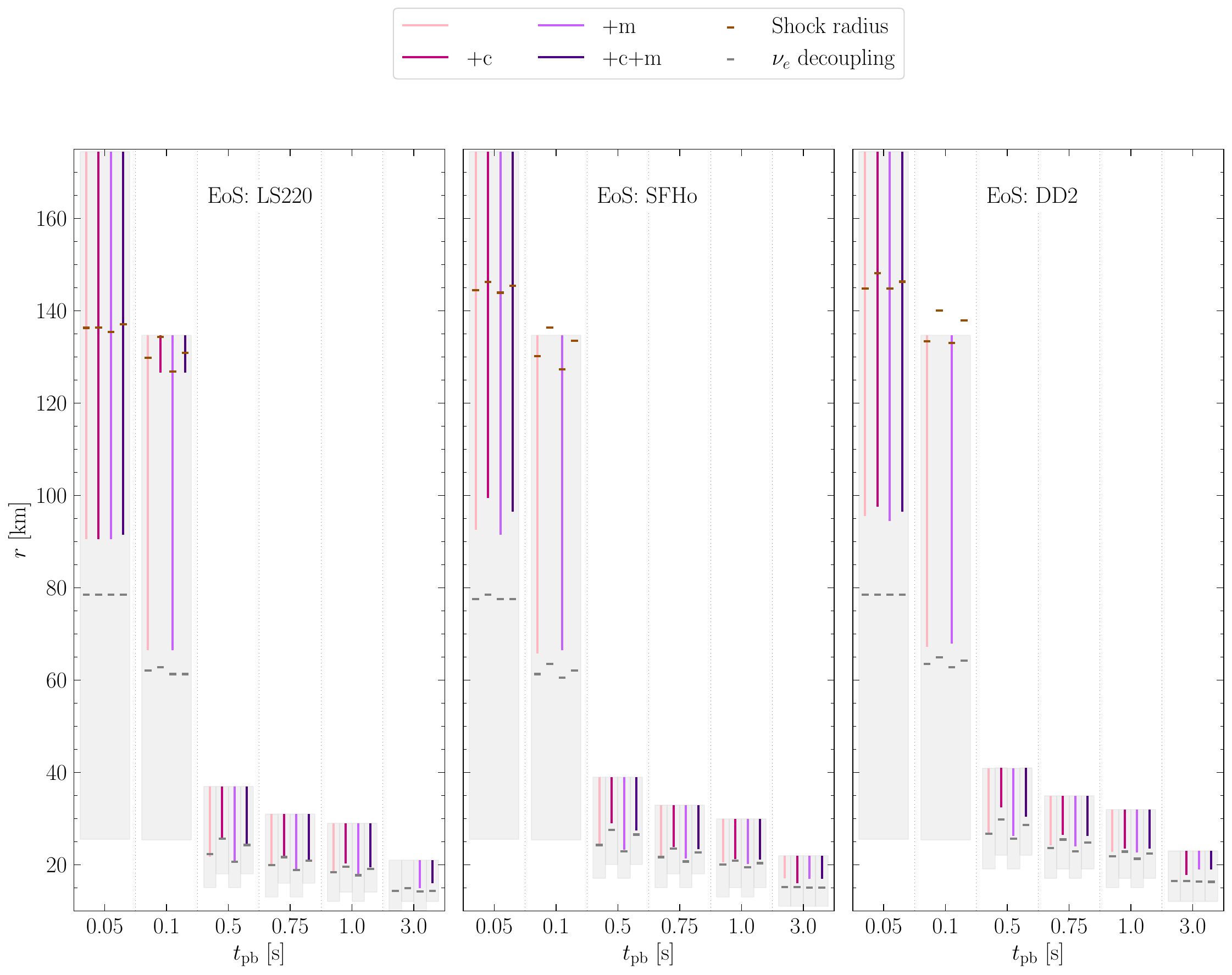}
    \caption{Radial range where ELN crossings are found in our suite of SN models for the selected post-bounce times (cf.~main text for details).
    The SN models are grouped by EoS: LS220 (left), SFHo (middle), and DD2 (right). The pink lines mark the radii at which an ELN crossing occurs for the SN models without muons and PNS convection, the magenta lines indicate ELN crossing loci for SN models with a mixing-length treatment of convection, the violet lines stand for SN models including muon production, and the purple lines represent the SN models with convection and muons. For orientation, for each post-bounce time, the locations of the shock radius and of the energy-averaged $\nu_e$ decoupling radius are plotted in brown and gray, respectively. The light gray bands highlight the radial range $[r_{\rm min}, r_{\rm max}]$ in which the Boltzmann equations are solved (and therefore the radial range in which we compute the angular distributions). Convection pushes ELN crossings to larger radii because it leads to larger PNS radii. On the other hand, muons facilitate the appearance of ELN crossings at smaller radii. These trends are more pronounced for the SFHo and DD2 EoSs.
    }
    \label{Fig:crossings_all_models}
\end{figure*}
Figure~\ref{Fig:crossings_all_models} provides a summary of the radial locations and post-bounce times where we find ELN crossings for $\cos\theta \gtrsim 0$ for all our SN models; note that, in order to obtain this figure and avoid spurious ELN crossings for $\cos\theta \gtrsim 0$, we use the selection criteria listed in Sec. III.A of Ref.~\cite{Cornelius:2025tyt}. Each panel of this figure groups SN models with the same EoS. For a specific SN model, the absence of a vertical line indicates that no crossing was found at the corresponding post-bounce time and radial bin. For orientation, the shock radius and the energy-averaged $\nu_e$ decoupling radius (defined as the radius where the $\nu_e$ flux factor $\simeq {1}/{3}$) are also plotted with brown and gray markers, respectively. 

We can see that ELN crossings exist for $\cos\theta \gtrsim 0$ at all post-bounce times, after neutrino decoupling. However, ELN crossings do not appear for the SN models including muons for $t_{\mathrm{pb}} = 0.1$~s (except for the LS220 EoS models near the upper radial boundary). ELN crossings are only present for the SN models with muons and without convection for $t_{\mathrm{pb}} = 3.0$~s for the LS220 EoS (although, we note that most of the crossings at 3 s are very weak, cf.~also Fig.~\ref{Fig:ELN}). 
The models with LS220 EoS for $t_{\rm pb} = 0.05$~s show smaller differences in the radial ranges where crossings occur than the ones with SFHo and DD2 EoS, while the differences between models become smaller for the remaining post-bounce times. 
Moreover, independent of the EoS, convection accelerates the PNS deleptonization, and we find that it tends to push the appearance of ELN crossings to larger radii. This is explained because crossings develop in the proximity of the minimum of the deleptonization trough in the $Y_e$ radial profile (see Fig.~\ref{Fig:hydro}). Such a dip is usually located around the $\nu_e$ neutrinosphere; convection has the effect of pushing the location of this feature to slightly larger radii. On the other hand, the presence of muons triggers a faster PNS contraction, with neutrino decoupling occurring at smaller radii, favoring the earlier appearance of ELN crossings.

\section{Discussion and outlook}
\label{sec:conclusions}
An accurate modeling of the neutrino angular distributions is essential to understand the flavor conversion physics in the SN core. Previous work explored the eventual appearance of ELN crossings in spherically symmetric and multi-dimensional hydrodynamic SN simulations~\cite{Tamborra:2017ubu,Nagakura:2021hyb,DelfanAzari:2019tez,Glas:2019ijo,Akaho:2022zdz,Akaho:2023brj,Nagakura:2019sig,Morinaga:2019wsv,Harada:2021ata,Abbar:2020qpi,Capozzi:2020syn}.
The main goal of this paper is to figure out whether the microphysics (e.g.,~the production of muons at high densities or the nuclear equation of state) or hydrodynamical effects (e.g.,~PNS convection) affect the formation of ELN crossings.

We rely on a set of $12$ spherically symmetric SN models with a progenitor mass of $18.6 M_\odot$. These models feature three nuclear EoSs: LS220, SFHo, and DD2. For fixed EoS, we consider a subset of SN models treating the effects of PNS convection by means of a mixing-length description, and another subset of models taking into account muon production in the SN core. 
We compute the neutrino angular distributions in the proximity of neutrino decoupling by solving the Boltzmann transport equations in post-processing. In order to do so, we use as input static fluid profiles extracted from the hydrodynamic SN simulations at selected post-bounce times. 
Except for a few instances, we find ELN crossings for $\cos\theta \gtrsim 0$ in all SN simulations from our suite, qualitatively confirming the trend presented in Refs.~\cite{Shalgar:2023aca,Shalgar:2024gjt}. 

Our results suggest that the location of the deleptonization dip of the electron fraction in the PNS convective layer is affected by convection and pushed to larger radii (because PNS convection leads to a slight increase of the PNS radius); as a consequence, ELN crossings tend to appear at larger radii. Since neutrino decoupling occurs at larger radii, the neutrino angular distributions appear broader in the presence of PNS convection. Note that Ref.~\cite{Nagakura:2021hyb} investigated the role of PNS convection on the existence of ELN crossings in multi-dimensional SN simulations and found that PNS convection causes both forward and backward crossings in the PNS convective layer, before neutrino decoupling. 
In this paper, we only focus on forward crossings and find that in our SN models--both with and without PNS convection--forward crossings appear after neutrino decoupling.

An opposite trend is instead observed when muon production is included in the SN simulations. The faster PNS contraction induced by muon creation facilitates neutrino decoupling, with ELN crossings appearing at smaller radii. 
It is worth noticing that, in the computation of the neutrino angular distributions, we do not distinguish between muon and tau (anti)neutrinos. However, beta processes involving muons could be responsible for differences between the $\mu$LN and $\tau$LN angular distributions and consequent crossings~\cite{Capozzi:2020syn}.

Qualitatively, we reach opposite conclusions with respect to the ones of Ref.~\cite{Tamborra:2017ubu} that did not find ELN crossings in any of their spherically symmetric models (with progenitor mass different than the $18.6 M_\odot$ one considered here, but with LS220 EoS, and in the absence of PNS convection and muon production). We have solved Eqs.~(\ref{Eq:KE}) relying on static fluid properties extracted from the SN simulations for the three post-bounce times shown in Fig.~6 of Ref.~\cite{Tamborra:2017ubu}: $0.256$~s (for the $11.2 M_\odot$ model, $[r_{\rm{min}}, r_{\rm{max}}] = [15, 55]$~km), $0.280$~s (for the $15 M_\odot$ model, $[r_{\rm{min}}, r_{\rm{max}}] = [16, 56]$~km), and $0.250$~s (for the $25 M_\odot$ model, $[r_{\rm{min}}, r_{\rm{max}}] = [20, 60]$~km). We find that, although the neutrino number densities for all flavors from the SN simulations and our Boltzmann solution agree up to $\mathcal{O}(10\%)$ (except near $r_{\rm{min}}$; see discussion in Appendix~\ref{appendix:comparison}), the fluxes show larger variations up to $\sim 50\%$ (results not shown here). Most likely these differences are due to the treatment of neutrino transport between our Boltzmann solution and the Boltzmann solver module used in the {\tt PROMETHEUS-VERTEX} neutrino-hydrodynamics code employed in Ref.~\cite{Tamborra:2017ubu}.
For example, in the setup employed here we consider a PNS atmosphere with boundary conditions that may not perfectly reproduce the neutrino properties at the location in the hydrodynamic model. 
Moreover, the collisional kernels employed in the two cases are somewhat different and may lead to subtle changes in the radial evolution of the neutrino fluxes and the $\nu_e$ and $\bar\nu_e$ angular distributions.
We have also tested in Appendix~\ref{appendix:comparison} that the energy resolution ($16$ bins up to $100$~MeV in Ref.~\cite{Tamborra:2017ubu} vs.~$100$ in our Boltzmann solution) leads to additional minor differences in the ELN angular distributions. 
Although we do not aim to reproduce with high fidelity the neutrino emission properties of the SN hydrodynamic simulations employing our Boltzmann solution (cf.~Appendix~\ref{appendix:comparison} for a comparison), our findings highlight the very subtle dependence of the ELN distribution on the microphysics and SN dynamics. 

Our analysis focused on spherically symmetric SN models. However, multi-dimensional hydrodynamical effects, such as LESA~\cite{Tamborra:2014aua}, favor the development of large-scale asymmetries that can further impact the ELN properties. We expect the overall dependence of the ELN angular distributions on microphysics and PNS convection explored in this work to be qualitatively similar in the context of multi-dimensional SN simulations, albeit new effects and direction-dependent differences can occur in multi-dimensional simulations.

We have focused our analysis on the (anti)neutrino angular distributions from SNe; however, our conclusions extend to neutron-star merger remnants where fast flavor instabilities are also expected to take place~\cite{Wu:2017qpc,Lund:2025jjo,Mukhopadhyay:2024zzl}. More work should be devoted to the conceptual development of new approaches facilitating the self-consistent modeling of neutrino transport including the angular distributions of (anti)neutrinos within hydrodynamical simulations.
Work in this direction will be crucial to robustly model the physics of flavor conversion in the core of neutrino-dense sources.

\section{Acknowledgments}
The authors thank Sajad Abbar, Daniel Kresse, and Georg Raffelt for useful discussions.
At the Niels Bohr Institute, this project has received support from the Villum Foundation (Project No.~13164) and the European Union (ERC, ANET, Project No.~101087058). 
In Garching, the work was supported by the German Research Foundation (DFG) through the Collaborative Research Centre ``Neutrinos and Dark Matter in Astro- and Particle Physics (NDM),'' Grant No.~SFB-1258-283604770, and under Germany's Excellence Strategy through the Cluster of Excellence ORIGINS EXC-2094-390783311. 
Views and opinions expressed are those of the authors only and do not necessarily reflect those of the European Union or the European Research Council. Neither the European Union nor the granting authority can be held responsible for them. The Tycho supercomputer hosted at the SCIENCE HPC Center at the University of Copenhagen was used to support the numerical simulations presented in this work. Computing resources are also acknowledged from the Max Planck Computing and Data Facility (MPCDF) on the HPC systems Cobra, Draco, and Raven. 

\appendix
\section{Comparison between the neutrino properties computed by solving the Boltzmann equations and the ones from the benchmark supernova model}
\label{appendix:comparison}
In this Appendix, we compare the first two angular moments extracted from our benchmark SN simulation with those computed relying on the Boltzmann equations. We also investigate how the energy resolution affects the neutrino angular distributions.

\begin{figure*}
    \centering
    \includegraphics[width=0.99\textwidth]{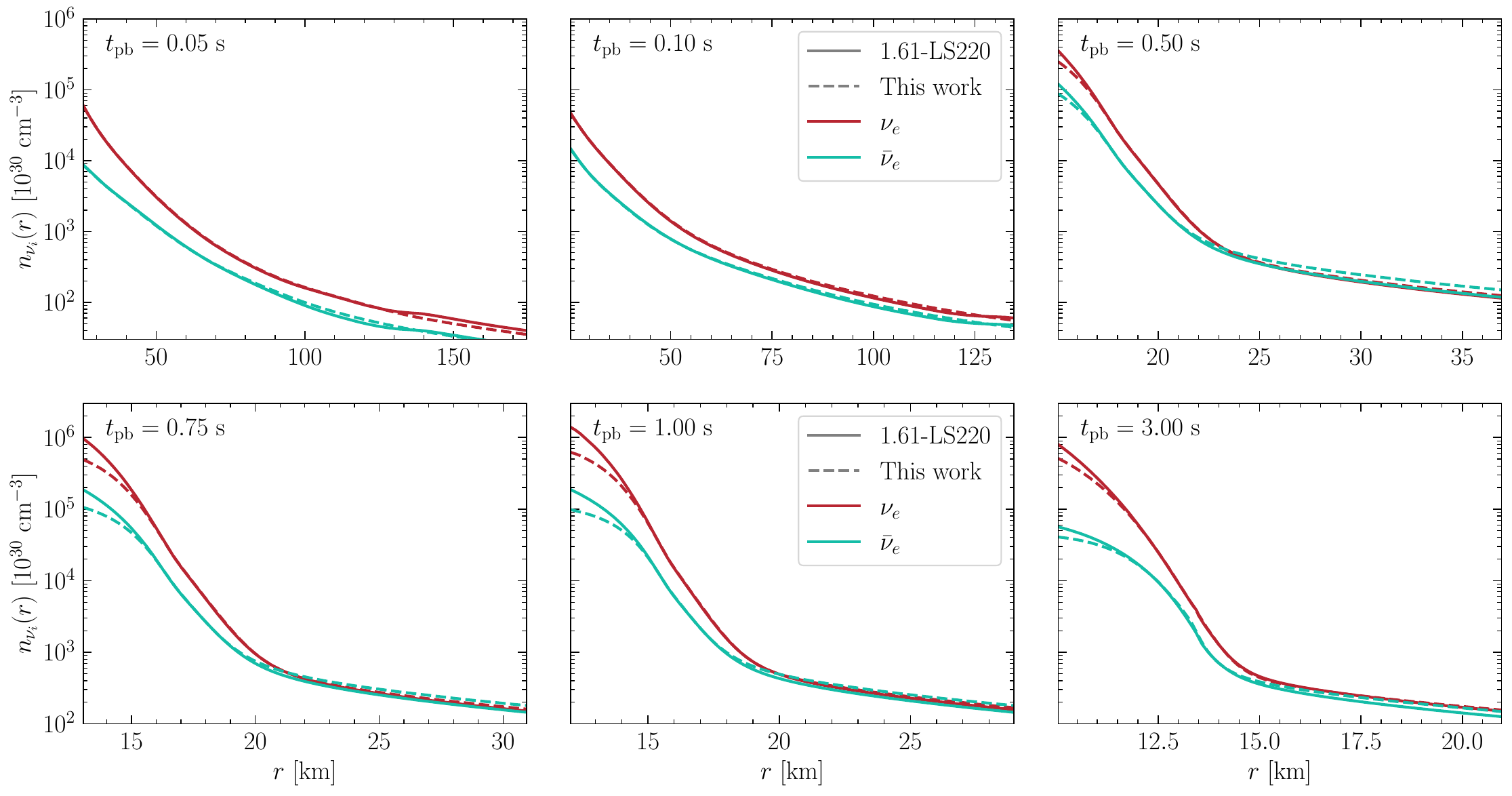}
    \caption{Radial profiles of the number densities of electron neutrinos (red) and antineutrinos (teal) for the benchmark SN model ($1.61$-LS220) and the solution of the Boltzmann equations between $r_{\rm{min}}$ and $r_{\rm{max}}$. The solid lines show the zeroth moment extracted from the SN simulation, while the dashed lines represent the number densities computed by solving the Boltzmann equations. Each panel represents one of the selected post-bounce times from $t_{\rm pb} = 0.05$~s (top left) to $t_{\rm pb} = 3$~s (bottom right). The number densities from the Boltzmann solution are in overall very good agreement with the ones from the SN simulation, with larger differences up to $\mathcal{O}(10)\%$ observable in the proximity of $r_{\rm min}$ for $t_{\rm pb} \gtrsim 0.75$~s and mostly affecting $\bar\nu_e$ at larger radii (see main text for details). 
   }
    \label{Fig:number_density}
\end{figure*}
The neutrino number density and number flux are given by
\begin{eqnarray}
  n_{\nu_\alpha}(r) &=& \frac{2\pi}{(hc)^3}\int \mathrm{d}E E^2 \int \mathrm{d}\cos\theta f_{\nu_\alpha}(r, \cos\theta, E)\ ,\\
  F_{\nu_\alpha}(r) &=& \frac{2\pi c}{(hc)^3}\int \mathrm{d}E E^2 \int \mathrm{d}\cos\theta~\cos\theta f_{\nu_\alpha}(r,\cos\theta,E) \ ,\nonumber\\
\end{eqnarray}
where $f_{\nu_\alpha}(r,\cos\theta,E)$ is the neutrino distribution function for the flavor $\nu_\alpha$.
Figure~\ref{Fig:number_density} shows the radial profiles of the electron neutrino and antineutrino number densities for our benchmark SN model ($1.61$-LS220) from the hydrodynamic SN simulation (solid lines) and our Boltzmann solution (dashed lines) for all six post-bounce times selected in this work. We can see that the neutrino number densities obtained from the Boltzmann solution differ from the ones extracted from the hydrodynamical simulation at most by few $\%$ for $\nu_e$'s, larger differences up to $\mathcal{O}(10)\%$ affect $\bar\nu_e$'s. 
The differences observable between the results of our Boltzmann equations and the output of the benchmark SN model become more prominent for $t_{\rm pb} \gtrsim 0.5$~s in the proximity of $r_{\rm min}$. This may be due to several factors. We solve the Boltzmann equations between $1$ and $100$~MeV, but the SN hydrodynamical simulations consider the energy range up to $380$~MeV. While the choice of solving the Boltzmann equations up to $100$~MeV allows us to use high-energy resolution without affecting the ELN crossing properties, the plasma temperature is higher in the SN core where neutrinos are coupled.
However, for $t_{\rm pb} \gtrsim 0.5$~s, the baryon density gradient in the proximity of $r_{\rm{min}}$ becomes steeper after the explosion has set in and PNS accretion has stopped. 
Moreover, because of the large baryon density in the proximity of $r_{\rm{min}}$, dense-matter effects in the collision integrals (e.g., nucleon recoil, nucleon correlations, and nuclear mean-field potentials) may become more relevant, but we do not account for this physics in our Boltzmann solution.
The larger differences in the $\bar\nu_e$ number densities are likely due to the different treatment of pair production between our Boltzmann solution and {\tt VERTEX}, and the fact that our Boltzmann solution does not account for inelastic scatterings for simplicity. 

\begin{figure*}
    \centerline{1.61-LS220}
    \vspace{0.5cm}
    \centering
    \includegraphics[width=0.99\textwidth]{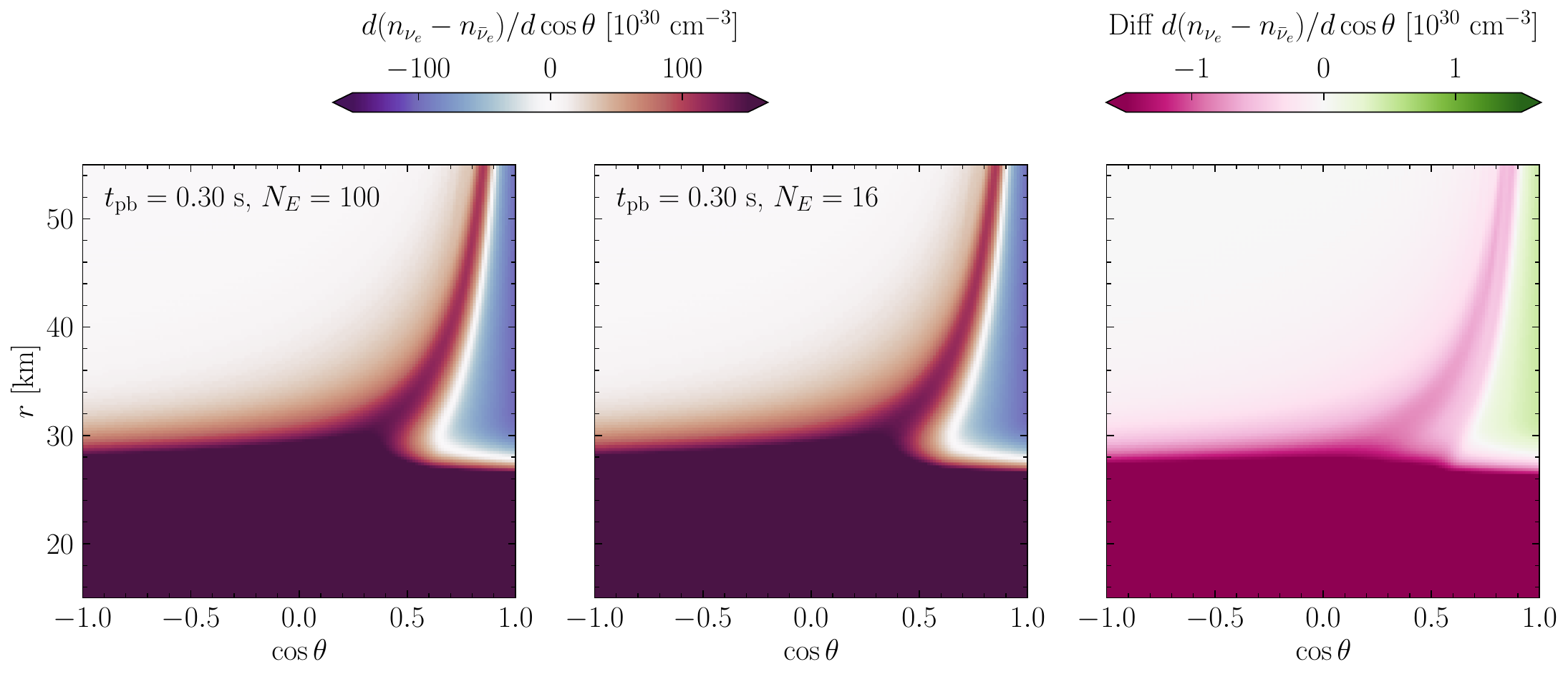}
    \caption{ELN differential number density for $t_{\rm pb} = 0.3$~s for the benchmark SN model ($1.61$-LS220) obtained adopting $100$ linearly spaced (left panel) and $16$ geometrically spaced (middle panel) bins between $0$ and $100$~MeV; the latter matches in our energy range the $21$ energy bins adopted in the hydrodynamical simulations that were run between $0$ and $380$~MeV. The right panel displays the difference between the ELN distributions with different energy resolutions.
    }
    \label{Fig:energy_resolution}
\end{figure*}
We also note that the $\bar\nu_e$ number density is always below the $\nu_e$ one in the radial profiles extracted from the SN simulations, as expected because the PNS accretion layer and contracting mantle deleptonize. However, our Boltzmann solutions show a flip between the two number densities towards larger radii for $t_{\rm pb}\gtrsim 0.5$~s. This effect was also pointed out in Ref.~\cite{Shalgar:2023aca} (cf.~their Appendix B) and it might affect the formation of ELN crossings. Nevertheless, we stress that it is beyond the scope of this paper to accurately reproduce the outputs of hydrodynamic simulations through an independent Boltzmann solution. Despite the differences in the implementation of the collisional kernel and neutrino transport, the agreement between our results is sufficiently good to use our Boltzmann calculations for exploring relative changes due to varied physics in the SN core. Moreover, relative changes of ELN crossing effects between the different cases with respect to the differences in the microphysics and fluid properties can be investigated sufficiently reliably by applying the Boltzmann code to given PNS profiles.

In order to investigate the dependence of our results on the energy resolution, Fig.~\ref{Fig:energy_resolution} displays the energy-integrated ELN number density obtained, for $1$ to $100$~MeV, with $100$ equidistant energy bins (left) and $16$ energy bins with the first six bins equidistant, then geometrically increasing (middle) as well as the differences between the differential $\nu_e$--$\bar\nu_e$ number densities computed in the two cases (right). Here, the number of energy bins matches the first $16$ out of $21$ bins employed in {\tt VERTEX} up to $100$~MeV.
We can see that the energy resolution mildly affects the appearance of ELN crossings. This finding contributes to strengthen our conclusions that the shape of the ELN angular distribution is sensitive to several aspects of the SN (micro)physics. It is important to compute the angular distributions of (anti)neutrinos with high accuracy since this can crucially affect the flavor outcome~\cite{Tamborra:2020cul,Padilla-Gay:2021haz}.

\newpage
\bibliography{references}

\begin{thebibliography}{67}%
\makeatletter
\providecommand \@ifxundefined [1]{%
 \@ifx{#1\undefined}
}%
\providecommand \@ifnum [1]{%
 \ifnum #1\expandafter \@firstoftwo
 \else \expandafter \@secondoftwo
 \fi
}%
\providecommand \@ifx [1]{%
 \ifx #1\expandafter \@firstoftwo
 \else \expandafter \@secondoftwo
 \fi
}%
\providecommand \natexlab [1]{#1}%
\providecommand \enquote  [1]{``#1''}%
\providecommand \bibnamefont  [1]{#1}%
\providecommand \bibfnamefont [1]{#1}%
\providecommand \citenamefont [1]{#1}%
\providecommand \href@noop [0]{\@secondoftwo}%
\providecommand \href [0]{\begingroup \@sanitize@url \@href}%
\providecommand \@href[1]{\@@startlink{#1}\@@href}%
\providecommand \@@href[1]{\endgroup#1\@@endlink}%
\providecommand \@sanitize@url [0]{\catcode `\\12\catcode `\$12\catcode
  `\&12\catcode `\#12\catcode `\^12\catcode `\_12\catcode `\%12\relax}%
\providecommand \@@startlink[1]{}%
\providecommand \@@endlink[0]{}%
\providecommand \url  [0]{\begingroup\@sanitize@url \@url }%
\providecommand \@url [1]{\endgroup\@href {#1}{\urlprefix }}%
\providecommand \urlprefix  [0]{URL }%
\providecommand \Eprint [0]{\href }%
\providecommand \doibase [0]{https://doi.org/}%
\providecommand \selectlanguage [0]{\@gobble}%
\providecommand \bibinfo  [0]{\@secondoftwo}%
\providecommand \bibfield  [0]{\@secondoftwo}%
\providecommand \translation [1]{[#1]}%
\providecommand \BibitemOpen [0]{}%
\providecommand \bibitemStop [0]{}%
\providecommand \bibitemNoStop [0]{.\EOS\space}%
\providecommand \EOS [0]{\spacefactor3000\relax}%
\providecommand \BibitemShut  [1]{\csname bibitem#1\endcsname}%
\let\auto@bib@innerbib\@empty
%</preamble>
\bibitem [{\citenamefont {Burrows}\ and\ \citenamefont
  {Vartanyan}(2021)}]{Burrows:2020qrp}%
  \BibitemOpen
  \bibfield  {author} {\bibinfo {author} {\bibfnamefont {A.}~\bibnamefont
  {Burrows}}\ and\ \bibinfo {author} {\bibfnamefont {D.}~\bibnamefont
  {Vartanyan}},\ }\bibfield  {title} {\bibinfo {title} {{Core-Collapse
  Supernova Explosion Theory}},\ }\href
  {https://doi.org/10.1038/s41586-020-03059-w} {\bibfield  {journal} {\bibinfo
  {journal} {Nature}\ }\textbf {\bibinfo {volume} {589}},\ \bibinfo {pages}
  {29} (\bibinfo {year} {2021})},\ \Eprint {https://arxiv.org/abs/2009.14157}
  {arXiv:2009.14157 [astro-ph.SR]} \BibitemShut {NoStop}%
\bibitem [{\citenamefont {Tamborra}(2025)}]{Tamborra:2024fcd}%
  \BibitemOpen
  \bibfield  {author} {\bibinfo {author} {\bibfnamefont {I.}~\bibnamefont
  {Tamborra}},\ }\bibfield  {title} {\bibinfo {title} {{Neutrinos from
  explosive transients at the dawn of multi-messenger astronomy}},\ }\href
  {https://doi.org/10.1038/s42254-025-00828-2} {\bibfield  {journal} {\bibinfo
  {journal} {Nature Rev. Phys.}\ }\textbf {\bibinfo {volume} {7}},\ \bibinfo
  {pages} {285} (\bibinfo {year} {2025})},\ \Eprint
  {https://arxiv.org/abs/2412.09699} {arXiv:2412.09699 [astro-ph.HE]}
  \BibitemShut {NoStop}%
\bibitem [{\citenamefont {Vitagliano}\ \emph {et~al.}(2020)\citenamefont
  {Vitagliano}, \citenamefont {Tamborra},\ and\ \citenamefont
  {Raffelt}}]{Vitagliano:2019yzm}%
  \BibitemOpen
  \bibfield  {author} {\bibinfo {author} {\bibfnamefont {E.}~\bibnamefont
  {Vitagliano}}, \bibinfo {author} {\bibfnamefont {I.}~\bibnamefont
  {Tamborra}},\ and\ \bibinfo {author} {\bibfnamefont {G.~G.}\ \bibnamefont
  {Raffelt}},\ }\bibfield  {title} {\bibinfo {title} {{Grand Unified Neutrino
  Spectrum at Earth: Sources and Spectral Components}},\ }\href
  {https://doi.org/10.1103/RevModPhys.92.045006} {\bibfield  {journal}
  {\bibinfo  {journal} {Rev. Mod. Phys.}\ }\textbf {\bibinfo {volume} {92}},\
  \bibinfo {pages} {045006} (\bibinfo {year} {2020})},\ \Eprint
  {https://arxiv.org/abs/1910.11878} {arXiv:1910.11878 [astro-ph.HE]}
  \BibitemShut {NoStop}%
\bibitem [{\citenamefont {Janka}(2025)}]{Janka:2025tvf}%
  \BibitemOpen
  \bibfield  {author} {\bibinfo {author} {\bibfnamefont {H.-T.}\ \bibnamefont
  {Janka}},\ }\href@noop {} {\bibinfo {title} {{Long-Term Multidimensional
  Models of Core-Collapse Supernovae: Progress and Challenges}}} (\bibinfo
  {year} {2025}),\ \Eprint {https://arxiv.org/abs/2502.14836} {arXiv:2502.14836
  [astro-ph.HE]} \BibitemShut {NoStop}%
\bibitem [{\citenamefont {Burrows}\ \emph {et~al.}(2018)\citenamefont
  {Burrows}, \citenamefont {Vartanyan}, \citenamefont {Dolence}, \citenamefont
  {Skinner},\ and\ \citenamefont {Radice}}]{Burrows:2016ohd}%
  \BibitemOpen
  \bibfield  {author} {\bibinfo {author} {\bibfnamefont {A.}~\bibnamefont
  {Burrows}}, \bibinfo {author} {\bibfnamefont {D.}~\bibnamefont {Vartanyan}},
  \bibinfo {author} {\bibfnamefont {J.~C.}\ \bibnamefont {Dolence}}, \bibinfo
  {author} {\bibfnamefont {M.~A.}\ \bibnamefont {Skinner}},\ and\ \bibinfo
  {author} {\bibfnamefont {D.}~\bibnamefont {Radice}},\ }\bibfield  {title}
  {\bibinfo {title} {{Crucial Physical Dependencies of the Core-Collapse
  Supernova Mechanism}},\ }\href {https://doi.org/10.1007/s11214-017-0450-9}
  {\bibfield  {journal} {\bibinfo  {journal} {Space Sci. Rev.}\ }\textbf
  {\bibinfo {volume} {214}},\ \bibinfo {pages} {33} (\bibinfo {year} {2018})},\
  \Eprint {https://arxiv.org/abs/1611.05859} {arXiv:1611.05859 [astro-ph.SR]}
  \BibitemShut {NoStop}%
\bibitem [{\citenamefont {Mezzacappa}\ \emph {et~al.}(2020)\citenamefont
  {Mezzacappa}, \citenamefont {Endeve}, \citenamefont {Bronson~Messer},\ and\
  \citenamefont {Bruenn}}]{Mezzacappa:2020oyq}%
  \BibitemOpen
  \bibfield  {author} {\bibinfo {author} {\bibfnamefont {A.}~\bibnamefont
  {Mezzacappa}}, \bibinfo {author} {\bibfnamefont {E.}~\bibnamefont {Endeve}},
  \bibinfo {author} {\bibfnamefont {O.~E.}\ \bibnamefont {Bronson~Messer}},\
  and\ \bibinfo {author} {\bibfnamefont {S.~W.}\ \bibnamefont {Bruenn}},\
  }\bibfield  {title} {\bibinfo {title} {{Physical, numerical, and
  computational challenges of modeling neutrino transport in core-collapse
  supernovae}},\ }\href {https://doi.org/10.1007/s41115-020-00010-8} {\bibfield
   {journal} {\bibinfo  {journal} {Liv. Rev. Comput. Astrophys.}\ }\textbf
  {\bibinfo {volume} {6}},\ \bibinfo {pages} {4} (\bibinfo {year} {2020})},\
  \Eprint {https://arxiv.org/abs/2010.09013} {arXiv:2010.09013 [astro-ph.HE]}
  \BibitemShut {NoStop}%
\bibitem [{\citenamefont {Tamborra}\ and\ \citenamefont
  {Shalgar}(2021)}]{Tamborra:2020cul}%
  \BibitemOpen
  \bibfield  {author} {\bibinfo {author} {\bibfnamefont {I.}~\bibnamefont
  {Tamborra}}\ and\ \bibinfo {author} {\bibfnamefont {S.}~\bibnamefont
  {Shalgar}},\ }\bibfield  {title} {\bibinfo {title} {{New Developments in
  Flavor Evolution of a Dense Neutrino Gas}},\ }\href
  {https://doi.org/10.1146/annurev-nucl-102920-050505} {\bibfield  {journal}
  {\bibinfo  {journal} {Ann. Rev. Nucl. Part. Sci.}\ }\textbf {\bibinfo
  {volume} {71}},\ \bibinfo {pages} {165} (\bibinfo {year} {2021})},\ \Eprint
  {https://arxiv.org/abs/2011.01948} {arXiv:2011.01948 [astro-ph.HE]}
  \BibitemShut {NoStop}%
\bibitem [{\citenamefont {Volpe}(2024)}]{Volpe:2023met}%
  \BibitemOpen
  \bibfield  {author} {\bibinfo {author} {\bibfnamefont {M.~C.}\ \bibnamefont
  {Volpe}},\ }\bibfield  {title} {\bibinfo {title} {{Neutrinos from dense
  environments: Flavor mechanisms, theoretical approaches, observations, and
  new directions}},\ }\href {https://doi.org/10.1103/RevModPhys.96.025004}
  {\bibfield  {journal} {\bibinfo  {journal} {Rev. Mod. Phys.}\ }\textbf
  {\bibinfo {volume} {96}},\ \bibinfo {pages} {025004} (\bibinfo {year}
  {2024})},\ \Eprint {https://arxiv.org/abs/2301.11814} {arXiv:2301.11814
  [hep-ph]} \BibitemShut {NoStop}%
\bibitem [{\citenamefont {Johns}\ \emph {et~al.}(2025)\citenamefont {Johns},
  \citenamefont {Richers},\ and\ \citenamefont {Wu}}]{Johns:2025mlm}%
  \BibitemOpen
  \bibfield  {author} {\bibinfo {author} {\bibfnamefont {L.}~\bibnamefont
  {Johns}}, \bibinfo {author} {\bibfnamefont {S.}~\bibnamefont {Richers}},\
  and\ \bibinfo {author} {\bibfnamefont {M.-R.}\ \bibnamefont {Wu}},\
  }\bibfield  {title} {\bibinfo {title} {{Neutrino Oscillations in
  Core-Collapse Supernovae and Neutron Star Mergers}},\ }\href
  {https://doi.org/10.1146/annurev-nucl-121423-100853} {\bibfield  {journal}
  {\bibinfo  {journal} {Ann. Rev. Nucl. Part. Sci.}\ }\textbf {\bibinfo
  {volume} {75}},\ \bibinfo {pages} {399} (\bibinfo {year} {2025})},\ \Eprint
  {https://arxiv.org/abs/2503.05959} {arXiv:2503.05959 [astro-ph.HE]}
  \BibitemShut {NoStop}%
\bibitem [{\citenamefont {Nagakura}(2023)}]{Nagakura:2023mhr}%
  \BibitemOpen
  \bibfield  {author} {\bibinfo {author} {\bibfnamefont {H.}~\bibnamefont
  {Nagakura}},\ }\bibfield  {title} {\bibinfo {title} {{Roles of Fast
  Neutrino-Flavor Conversion on the Neutrino-Heating Mechanism of Core-Collapse
  Supernova}},\ }\href {https://doi.org/10.1103/PhysRevLett.130.211401}
  {\bibfield  {journal} {\bibinfo  {journal} {Phys. Rev. Lett.}\ }\textbf
  {\bibinfo {volume} {130}},\ \bibinfo {pages} {211401} (\bibinfo {year}
  {2023})},\ \Eprint {https://arxiv.org/abs/2301.10785} {arXiv:2301.10785
  [astro-ph.HE]} \BibitemShut {NoStop}%
\bibitem [{\citenamefont {Ehring}\ \emph
  {et~al.}(2023{\natexlab{a}})\citenamefont {Ehring}, \citenamefont {Abbar},
  \citenamefont {Janka}, \citenamefont {Raffelt},\ and\ \citenamefont
  {Tamborra}}]{Ehring:2023abs}%
  \BibitemOpen
  \bibfield  {author} {\bibinfo {author} {\bibfnamefont {J.}~\bibnamefont
  {Ehring}}, \bibinfo {author} {\bibfnamefont {S.}~\bibnamefont {Abbar}},
  \bibinfo {author} {\bibfnamefont {H.-T.}\ \bibnamefont {Janka}}, \bibinfo
  {author} {\bibfnamefont {G.~G.}\ \bibnamefont {Raffelt}},\ and\ \bibinfo
  {author} {\bibfnamefont {I.}~\bibnamefont {Tamborra}},\ }\bibfield  {title}
  {\bibinfo {title} {{Fast Neutrino Flavor Conversions Can Help and Hinder
  Neutrino-Driven Explosions}},\ }\href
  {https://doi.org/10.1103/PhysRevLett.131.061401} {\bibfield  {journal}
  {\bibinfo  {journal} {Phys. Rev. Lett.}\ }\textbf {\bibinfo {volume} {131}},\
  \bibinfo {pages} {061401} (\bibinfo {year} {2023}{\natexlab{a}})},\ \Eprint
  {https://arxiv.org/abs/2305.11207} {arXiv:2305.11207 [astro-ph.HE]}
  \BibitemShut {NoStop}%
\bibitem [{\citenamefont {Ehring}\ \emph
  {et~al.}(2023{\natexlab{b}})\citenamefont {Ehring}, \citenamefont {Abbar},
  \citenamefont {Janka}, \citenamefont {Raffelt},\ and\ \citenamefont
  {Tamborra}}]{Ehring:2023lcd}%
  \BibitemOpen
  \bibfield  {author} {\bibinfo {author} {\bibfnamefont {J.}~\bibnamefont
  {Ehring}}, \bibinfo {author} {\bibfnamefont {S.}~\bibnamefont {Abbar}},
  \bibinfo {author} {\bibfnamefont {H.-T.}\ \bibnamefont {Janka}}, \bibinfo
  {author} {\bibfnamefont {G.~G.}\ \bibnamefont {Raffelt}},\ and\ \bibinfo
  {author} {\bibfnamefont {I.}~\bibnamefont {Tamborra}},\ }\bibfield  {title}
  {\bibinfo {title} {{Fast neutrino flavor conversion in core-collapse
  supernovae: A parametric study in 1D models}},\ }\href
  {https://doi.org/10.1103/PhysRevD.107.103034} {\bibfield  {journal} {\bibinfo
   {journal} {Phys. Rev. D}\ }\textbf {\bibinfo {volume} {107}},\ \bibinfo
  {pages} {103034} (\bibinfo {year} {2023}{\natexlab{b}})},\ \Eprint
  {https://arxiv.org/abs/2301.11938} {arXiv:2301.11938 [astro-ph.HE]}
  \BibitemShut {NoStop}%
\bibitem [{\citenamefont {Mori}\ \emph {et~al.}(2025)\citenamefont {Mori},
  \citenamefont {Takiwaki}, \citenamefont {Kotake},\ and\ \citenamefont
  {Horiuchi}}]{Mori:2025cke}%
  \BibitemOpen
  \bibfield  {author} {\bibinfo {author} {\bibfnamefont {K.}~\bibnamefont
  {Mori}}, \bibinfo {author} {\bibfnamefont {T.}~\bibnamefont {Takiwaki}},
  \bibinfo {author} {\bibfnamefont {K.}~\bibnamefont {Kotake}},\ and\ \bibinfo
  {author} {\bibfnamefont {S.}~\bibnamefont {Horiuchi}},\ }\bibfield  {title}
  {\bibinfo {title} {{Three-dimensional core-collapse supernova models with
  phenomenological treatment of neutrino flavor conversions}},\ }\href
  {https://doi.org/10.1093/pasj/psaf007} {\bibfield  {journal} {\bibinfo
  {journal} {Publ. Astron. Soc. Jap.}\ }\textbf {\bibinfo {volume} {77}},\
  \bibinfo {pages} {L9} (\bibinfo {year} {2025})},\ \Eprint
  {https://arxiv.org/abs/2501.15256} {arXiv:2501.15256 [astro-ph.HE]}
  \BibitemShut {NoStop}%
\bibitem [{\citenamefont {{Wang}}\ and\ \citenamefont
  {{Burrows}}(2025)}]{Wang:2025nii}%
  \BibitemOpen
  \bibfield  {author} {\bibinfo {author} {\bibfnamefont {T.}~\bibnamefont
  {{Wang}}}\ and\ \bibinfo {author} {\bibfnamefont {A.}~\bibnamefont
  {{Burrows}}},\ }\bibfield  {title} {\bibinfo {title} {{The Effect of the
  Fast-flavor Instability on Core-collapse Supernova Models}},\ }\href
  {https://doi.org/10.3847/1538-4357/add889} {\bibfield  {journal} {\bibinfo
  {journal} {Astrophys. J.}\ }\textbf {\bibinfo {volume} {986}},\ \bibinfo
  {eid} {153} (\bibinfo {year} {2025})},\ \Eprint
  {https://arxiv.org/abs/2503.04896} {arXiv:2503.04896 [astro-ph.HE]}
  \BibitemShut {NoStop}%
\bibitem [{\citenamefont {Sawyer}(2005)}]{Sawyer:2005jk}%
  \BibitemOpen
  \bibfield  {author} {\bibinfo {author} {\bibfnamefont {R.~F.}\ \bibnamefont
  {Sawyer}},\ }\bibfield  {title} {\bibinfo {title} {{Speed-up of neutrino
  transformations in a supernova environment}},\ }\href
  {https://doi.org/10.1103/PhysRevD.72.045003} {\bibfield  {journal} {\bibinfo
  {journal} {Phys. Rev. D}\ }\textbf {\bibinfo {volume} {72}},\ \bibinfo
  {pages} {045003} (\bibinfo {year} {2005})},\ \Eprint
  {https://arxiv.org/abs/hep-ph/0503013} {arXiv:hep-ph/0503013} \BibitemShut
  {NoStop}%
\bibitem [{\citenamefont {Sawyer}(2016)}]{Sawyer:2015dsa}%
  \BibitemOpen
  \bibfield  {author} {\bibinfo {author} {\bibfnamefont {R.~F.}\ \bibnamefont
  {Sawyer}},\ }\bibfield  {title} {\bibinfo {title} {{Neutrino cloud
  instabilities just above the neutrino sphere of a supernova}},\ }\href
  {https://doi.org/10.1103/PhysRevLett.116.081101} {\bibfield  {journal}
  {\bibinfo  {journal} {Phys. Rev. Lett.}\ }\textbf {\bibinfo {volume} {116}},\
  \bibinfo {pages} {081101} (\bibinfo {year} {2016})},\ \Eprint
  {https://arxiv.org/abs/1509.03323} {arXiv:1509.03323 [astro-ph.HE]}
  \BibitemShut {NoStop}%
\bibitem [{\citenamefont {Chakraborty}\ \emph
  {et~al.}(2016{\natexlab{a}})\citenamefont {Chakraborty}, \citenamefont
  {Hansen}, \citenamefont {Izaguirre},\ and\ \citenamefont
  {Raffelt}}]{Chakraborty:2016yeg}%
  \BibitemOpen
  \bibfield  {author} {\bibinfo {author} {\bibfnamefont {S.}~\bibnamefont
  {Chakraborty}}, \bibinfo {author} {\bibfnamefont {R.}~\bibnamefont {Hansen}},
  \bibinfo {author} {\bibfnamefont {I.}~\bibnamefont {Izaguirre}},\ and\
  \bibinfo {author} {\bibfnamefont {G.~G.}\ \bibnamefont {Raffelt}},\
  }\bibfield  {title} {\bibinfo {title} {{Collective neutrino flavor
  conversion: Recent developments}},\ }\href
  {https://doi.org/10.1016/j.nuclphysb.2016.02.012} {\bibfield  {journal}
  {\bibinfo  {journal} {Nucl. Phys. B}\ }\textbf {\bibinfo {volume} {908}},\
  \bibinfo {pages} {366} (\bibinfo {year} {2016}{\natexlab{a}})},\ \Eprint
  {https://arxiv.org/abs/1602.02766} {arXiv:1602.02766 [hep-ph]} \BibitemShut
  {NoStop}%
\bibitem [{\citenamefont {Chakraborty}\ \emph
  {et~al.}(2016{\natexlab{b}})\citenamefont {Chakraborty}, \citenamefont
  {Hansen}, \citenamefont {Izaguirre},\ and\ \citenamefont
  {Raffelt}}]{Chakraborty:2016lct}%
  \BibitemOpen
  \bibfield  {author} {\bibinfo {author} {\bibfnamefont {S.}~\bibnamefont
  {Chakraborty}}, \bibinfo {author} {\bibfnamefont {R.~S.}\ \bibnamefont
  {Hansen}}, \bibinfo {author} {\bibfnamefont {I.}~\bibnamefont {Izaguirre}},\
  and\ \bibinfo {author} {\bibfnamefont {G.~G.}\ \bibnamefont {Raffelt}},\
  }\bibfield  {title} {\bibinfo {title} {{Self-induced neutrino flavor
  conversion without flavor mixing}},\ }\href
  {https://doi.org/10.1088/1475-7516/2016/03/042} {\bibfield  {journal}
  {\bibinfo  {journal} {JCAP}\ }\textbf {\bibinfo {volume} {03}},\ \bibinfo
  {pages} {042}},\ \Eprint {https://arxiv.org/abs/1602.00698} {arXiv:1602.00698
  [hep-ph]} \BibitemShut {NoStop}%
\bibitem [{\citenamefont {Izaguirre}\ \emph {et~al.}(2017)\citenamefont
  {Izaguirre}, \citenamefont {Raffelt},\ and\ \citenamefont
  {Tamborra}}]{Izaguirre:2016gsx}%
  \BibitemOpen
  \bibfield  {author} {\bibinfo {author} {\bibfnamefont {I.}~\bibnamefont
  {Izaguirre}}, \bibinfo {author} {\bibfnamefont {G.~G.}\ \bibnamefont
  {Raffelt}},\ and\ \bibinfo {author} {\bibfnamefont {I.}~\bibnamefont
  {Tamborra}},\ }\bibfield  {title} {\bibinfo {title} {{Fast Pairwise
  Conversion of Supernova Neutrinos: A Dispersion-Relation Approach}},\ }\href
  {https://doi.org/10.1103/PhysRevLett.118.021101} {\bibfield  {journal}
  {\bibinfo  {journal} {Phys. Rev. Lett.}\ }\textbf {\bibinfo {volume} {118}},\
  \bibinfo {pages} {021101} (\bibinfo {year} {2017})},\ \Eprint
  {https://arxiv.org/abs/1610.01612} {arXiv:1610.01612 [hep-ph]} \BibitemShut
  {NoStop}%
\bibitem [{\citenamefont {Morinaga}(2022)}]{Morinaga:2021vmc}%
  \BibitemOpen
  \bibfield  {author} {\bibinfo {author} {\bibfnamefont {T.}~\bibnamefont
  {Morinaga}},\ }\bibfield  {title} {\bibinfo {title} {{Fast neutrino flavor
  instability and neutrino flavor lepton number crossings}},\ }\href
  {https://doi.org/10.1103/PhysRevD.105.L101301} {\bibfield  {journal}
  {\bibinfo  {journal} {Phys. Rev. D}\ }\textbf {\bibinfo {volume} {105}},\
  \bibinfo {pages} {L101301} (\bibinfo {year} {2022})},\ \Eprint
  {https://arxiv.org/abs/2103.15267} {arXiv:2103.15267 [hep-ph]} \BibitemShut
  {NoStop}%
\bibitem [{\citenamefont {Padilla-Gay}\ \emph {et~al.}(2022)\citenamefont
  {Padilla-Gay}, \citenamefont {Tamborra},\ and\ \citenamefont
  {Raffelt}}]{Padilla-Gay:2021haz}%
  \BibitemOpen
  \bibfield  {author} {\bibinfo {author} {\bibfnamefont {I.}~\bibnamefont
  {Padilla-Gay}}, \bibinfo {author} {\bibfnamefont {I.}~\bibnamefont
  {Tamborra}},\ and\ \bibinfo {author} {\bibfnamefont {G.~G.}\ \bibnamefont
  {Raffelt}},\ }\bibfield  {title} {\bibinfo {title} {{Neutrino Flavor Pendulum
  Reloaded: The Case of Fast Pairwise Conversion}},\ }\href
  {https://doi.org/10.1103/PhysRevLett.128.121102} {\bibfield  {journal}
  {\bibinfo  {journal} {Phys. Rev. Lett.}\ }\textbf {\bibinfo {volume} {128}},\
  \bibinfo {pages} {121102} (\bibinfo {year} {2022})},\ \Eprint
  {https://arxiv.org/abs/2109.14627} {arXiv:2109.14627 [astro-ph.HE]}
  \BibitemShut {NoStop}%
\bibitem [{\citenamefont {{Fiorillo}}\ and\ \citenamefont
  {{Raffelt}}(2023)}]{2023PhRvD.107l3024F}%
  \BibitemOpen
  \bibfield  {author} {\bibinfo {author} {\bibfnamefont {D.~F.~G.}\
  \bibnamefont {{Fiorillo}}}\ and\ \bibinfo {author} {\bibfnamefont {G.~G.}\
  \bibnamefont {{Raffelt}}},\ }\bibfield  {title} {\bibinfo {title} {{Flavor
  solitons in dense neutrino gases}},\ }\href
  {https://doi.org/10.1103/PhysRevD.107.123024} {\bibfield  {journal} {\bibinfo
   {journal} {Phys. Rev. D}\ }\textbf {\bibinfo {volume} {107}},\ \bibinfo
  {eid} {123024} (\bibinfo {year} {2023})},\ \Eprint
  {https://arxiv.org/abs/2303.12143} {arXiv:2303.12143 [hep-ph]} \BibitemShut
  {NoStop}%
\bibitem [{\citenamefont {Tamborra}\ \emph {et~al.}(2017)\citenamefont
  {Tamborra}, \citenamefont {H{\"u}depohl}, \citenamefont {Raffelt},\ and\
  \citenamefont {Janka}}]{Tamborra:2017ubu}%
  \BibitemOpen
  \bibfield  {author} {\bibinfo {author} {\bibfnamefont {I.}~\bibnamefont
  {Tamborra}}, \bibinfo {author} {\bibfnamefont {L.}~\bibnamefont
  {H{\"u}depohl}}, \bibinfo {author} {\bibfnamefont {G.~G.}\ \bibnamefont
  {Raffelt}},\ and\ \bibinfo {author} {\bibfnamefont {H.-T.}\ \bibnamefont
  {Janka}},\ }\bibfield  {title} {\bibinfo {title} {{Flavor-dependent neutrino
  angular distribution in core-collapse supernovae}},\ }\href
  {https://doi.org/10.3847/1538-4357/aa6a18} {\bibfield  {journal} {\bibinfo
  {journal} {Astrophys. J.}\ }\textbf {\bibinfo {volume} {839}},\ \bibinfo
  {pages} {132} (\bibinfo {year} {2017})},\ \Eprint
  {https://arxiv.org/abs/1702.00060} {arXiv:1702.00060 [astro-ph.HE]}
  \BibitemShut {NoStop}%
\bibitem [{\citenamefont {Brandt}\ \emph {et~al.}(2011)\citenamefont {Brandt},
  \citenamefont {Burrows}, \citenamefont {Ott},\ and\ \citenamefont
  {Livne}}]{Brandt:2010xa}%
  \BibitemOpen
  \bibfield  {author} {\bibinfo {author} {\bibfnamefont {T.~D.}\ \bibnamefont
  {Brandt}}, \bibinfo {author} {\bibfnamefont {A.}~\bibnamefont {Burrows}},
  \bibinfo {author} {\bibfnamefont {C.~D.}\ \bibnamefont {Ott}},\ and\ \bibinfo
  {author} {\bibfnamefont {E.}~\bibnamefont {Livne}},\ }\bibfield  {title}
  {\bibinfo {title} {{Results From Core-Collapse Simulations with
  Multi-Dimensional, Multi-Angle Neutrino Transport}},\ }\href
  {https://doi.org/10.1088/0004-637X/728/1/8} {\bibfield  {journal} {\bibinfo
  {journal} {Astrophys. J.}\ }\textbf {\bibinfo {volume} {728}},\ \bibinfo
  {pages} {8} (\bibinfo {year} {2011})},\ \Eprint
  {https://arxiv.org/abs/1009.4654} {arXiv:1009.4654 [astro-ph.HE]}
  \BibitemShut {NoStop}%
\bibitem [{\citenamefont {Shalgar}\ and\ \citenamefont
  {Tamborra}(2019)}]{Shalgar:2019kzy}%
  \BibitemOpen
  \bibfield  {author} {\bibinfo {author} {\bibfnamefont {S.}~\bibnamefont
  {Shalgar}}\ and\ \bibinfo {author} {\bibfnamefont {I.}~\bibnamefont
  {Tamborra}},\ }\bibfield  {title} {\bibinfo {title} {{On the Occurrence of
  Crossings Between the Angular Distributions of Electron Neutrinos and
  Antineutrinos in the Supernova Core}},\ }\href
  {https://doi.org/10.3847/1538-4357/ab38ba} {\bibfield  {journal} {\bibinfo
  {journal} {Astrophys. J.}\ }\textbf {\bibinfo {volume} {883}},\ \bibinfo
  {pages} {80} (\bibinfo {year} {2019})},\ \Eprint
  {https://arxiv.org/abs/1904.07236} {arXiv:1904.07236 [astro-ph.HE]}
  \BibitemShut {NoStop}%
\bibitem [{\citenamefont {Sumiyoshi}\ and\ \citenamefont
  {Yamada}(2012)}]{Sumiyoshi:2012za}%
  \BibitemOpen
  \bibfield  {author} {\bibinfo {author} {\bibfnamefont {K.}~\bibnamefont
  {Sumiyoshi}}\ and\ \bibinfo {author} {\bibfnamefont {S.}~\bibnamefont
  {Yamada}},\ }\bibfield  {title} {\bibinfo {title} {{Neutrino Transfer in
  Three Dimensions for Core-Collapse Supernovae. I. Static Configurations}},\
  }\href {https://doi.org/10.1088/0067-0049/199/1/17} {\bibfield  {journal}
  {\bibinfo  {journal} {Astrophys. J. Suppl.}\ }\textbf {\bibinfo {volume}
  {199}},\ \bibinfo {pages} {17} (\bibinfo {year} {2012})},\ \Eprint
  {https://arxiv.org/abs/1201.2244} {arXiv:1201.2244 [astro-ph.HE]}
  \BibitemShut {NoStop}%
\bibitem [{\citenamefont {Akaho}\ \emph {et~al.}(2021)\citenamefont {Akaho},
  \citenamefont {Harada}, \citenamefont {Nagakura}, \citenamefont {Sumiyoshi},
  \citenamefont {Iwakami}, \citenamefont {Okawa}, \citenamefont {Furusawa},
  \citenamefont {Matsufuru},\ and\ \citenamefont {Yamada}}]{Akaho:2020xgb}%
  \BibitemOpen
  \bibfield  {author} {\bibinfo {author} {\bibfnamefont {R.}~\bibnamefont
  {Akaho}}, \bibinfo {author} {\bibfnamefont {A.}~\bibnamefont {Harada}},
  \bibinfo {author} {\bibfnamefont {H.}~\bibnamefont {Nagakura}}, \bibinfo
  {author} {\bibfnamefont {K.}~\bibnamefont {Sumiyoshi}}, \bibinfo {author}
  {\bibfnamefont {W.}~\bibnamefont {Iwakami}}, \bibinfo {author} {\bibfnamefont
  {H.}~\bibnamefont {Okawa}}, \bibinfo {author} {\bibfnamefont
  {S.}~\bibnamefont {Furusawa}}, \bibinfo {author} {\bibfnamefont
  {H.}~\bibnamefont {Matsufuru}},\ and\ \bibinfo {author} {\bibfnamefont
  {S.}~\bibnamefont {Yamada}},\ }\bibfield  {title} {\bibinfo {title}
  {{Multidimensional Boltzmann Neutrino Transport Code in Full General
  Relativity for Core-collapse Simulations}},\ }\href
  {https://doi.org/10.3847/1538-4357/abe1bf} {\bibfield  {journal} {\bibinfo
  {journal} {Astrophys. J.}\ }\textbf {\bibinfo {volume} {909}},\ \bibinfo
  {pages} {210} (\bibinfo {year} {2021})},\ \Eprint
  {https://arxiv.org/abs/2010.10780} {arXiv:2010.10780 [astro-ph.HE]}
  \BibitemShut {NoStop}%
\bibitem [{\citenamefont {Nagakura}\ \emph {et~al.}(2018)\citenamefont
  {Nagakura}, \citenamefont {Iwakami}, \citenamefont {Furusawa}, \citenamefont
  {Okawa}, \citenamefont {Harada}, \citenamefont {Sumiyoshi}, \citenamefont
  {Yamada}, \citenamefont {Matsufuru},\ and\ \citenamefont
  {Imakura}}]{Nagakura:2017mnp}%
  \BibitemOpen
  \bibfield  {author} {\bibinfo {author} {\bibfnamefont {H.}~\bibnamefont
  {Nagakura}}, \bibinfo {author} {\bibfnamefont {W.}~\bibnamefont {Iwakami}},
  \bibinfo {author} {\bibfnamefont {S.}~\bibnamefont {Furusawa}}, \bibinfo
  {author} {\bibfnamefont {H.}~\bibnamefont {Okawa}}, \bibinfo {author}
  {\bibfnamefont {A.}~\bibnamefont {Harada}}, \bibinfo {author} {\bibfnamefont
  {K.}~\bibnamefont {Sumiyoshi}}, \bibinfo {author} {\bibfnamefont
  {S.}~\bibnamefont {Yamada}}, \bibinfo {author} {\bibfnamefont
  {H.}~\bibnamefont {Matsufuru}},\ and\ \bibinfo {author} {\bibfnamefont
  {A.}~\bibnamefont {Imakura}},\ }\bibfield  {title} {\bibinfo {title}
  {{Simulations of core-collapse supernovae in spatial axisymmetry with full
  Boltzmann neutrino transport}},\ }\href
  {https://doi.org/10.3847/1538-4357/aaac29} {\bibfield  {journal} {\bibinfo
  {journal} {Astrophys. J.}\ }\textbf {\bibinfo {volume} {854}},\ \bibinfo
  {pages} {136} (\bibinfo {year} {2018})},\ \Eprint
  {https://arxiv.org/abs/1702.01752} {arXiv:1702.01752 [astro-ph.HE]}
  \BibitemShut {NoStop}%
\bibitem [{\citenamefont {Fischer}\ \emph {et~al.}(2010)\citenamefont
  {Fischer}, \citenamefont {Whitehouse}, \citenamefont {Mezzacappa},
  \citenamefont {Thielemann},\ and\ \citenamefont
  {Liebend{\"o}rfer}}]{Fischer:2009af}%
  \BibitemOpen
  \bibfield  {author} {\bibinfo {author} {\bibfnamefont {T.}~\bibnamefont
  {Fischer}}, \bibinfo {author} {\bibfnamefont {S.~C.}\ \bibnamefont
  {Whitehouse}}, \bibinfo {author} {\bibfnamefont {A.}~\bibnamefont
  {Mezzacappa}}, \bibinfo {author} {\bibfnamefont {F.~K.}\ \bibnamefont
  {Thielemann}},\ and\ \bibinfo {author} {\bibfnamefont {M.}~\bibnamefont
  {Liebend{\"o}rfer}},\ }\bibfield  {title} {\bibinfo {title} {{Protoneutron
  star evolution and the neutrino driven wind in general relativistic neutrino
  radiation hydrodynamics simulations}},\ }\href
  {https://doi.org/10.1051/0004-6361/200913106} {\bibfield  {journal} {\bibinfo
   {journal} {Astron. Astrophys.}\ }\textbf {\bibinfo {volume} {517}},\
  \bibinfo {pages} {A80} (\bibinfo {year} {2010})},\ \Eprint
  {https://arxiv.org/abs/0908.1871} {arXiv:0908.1871 [astro-ph.HE]}
  \BibitemShut {NoStop}%
\bibitem [{\citenamefont {Abbar}(2020)}]{Abbar:2020fcl}%
  \BibitemOpen
  \bibfield  {author} {\bibinfo {author} {\bibfnamefont {S.}~\bibnamefont
  {Abbar}},\ }\bibfield  {title} {\bibinfo {title} {{Searching for Fast
  Neutrino Flavor Conversion Modes in Core-collapse Supernova Simulations}},\
  }\href {https://doi.org/10.1088/1475-7516/2020/05/027} {\bibfield  {journal}
  {\bibinfo  {journal} {JCAP}\ }\textbf {\bibinfo {volume} {05}},\ \bibinfo
  {pages} {027}},\ \Eprint {https://arxiv.org/abs/2003.00969} {arXiv:2003.00969
  [astro-ph.HE]} \BibitemShut {NoStop}%
\bibitem [{\citenamefont {Abbar}\ \emph {et~al.}(2021)\citenamefont {Abbar},
  \citenamefont {Capozzi}, \citenamefont {Glas}, \citenamefont {Janka},\ and\
  \citenamefont {Tamborra}}]{Abbar:2020qpi}%
  \BibitemOpen
  \bibfield  {author} {\bibinfo {author} {\bibfnamefont {S.}~\bibnamefont
  {Abbar}}, \bibinfo {author} {\bibfnamefont {F.}~\bibnamefont {Capozzi}},
  \bibinfo {author} {\bibfnamefont {R.}~\bibnamefont {Glas}}, \bibinfo {author}
  {\bibfnamefont {H.-T.}\ \bibnamefont {Janka}},\ and\ \bibinfo {author}
  {\bibfnamefont {I.}~\bibnamefont {Tamborra}},\ }\bibfield  {title} {\bibinfo
  {title} {{On the characteristics of fast neutrino flavor instabilities in
  three-dimensional core-collapse supernova models}},\ }\href
  {https://doi.org/10.1103/PhysRevD.103.063033} {\bibfield  {journal} {\bibinfo
   {journal} {Phys. Rev. D}\ }\textbf {\bibinfo {volume} {103}},\ \bibinfo
  {pages} {063033} (\bibinfo {year} {2021})},\ \Eprint
  {https://arxiv.org/abs/2012.06594} {arXiv:2012.06594 [astro-ph.HE]}
  \BibitemShut {NoStop}%
\bibitem [{\citenamefont {Capozzi}\ \emph {et~al.}(2021)\citenamefont
  {Capozzi}, \citenamefont {Abbar}, \citenamefont {Bollig},\ and\ \citenamefont
  {Janka}}]{Capozzi:2020syn}%
  \BibitemOpen
  \bibfield  {author} {\bibinfo {author} {\bibfnamefont {F.}~\bibnamefont
  {Capozzi}}, \bibinfo {author} {\bibfnamefont {S.}~\bibnamefont {Abbar}},
  \bibinfo {author} {\bibfnamefont {R.}~\bibnamefont {Bollig}},\ and\ \bibinfo
  {author} {\bibfnamefont {H.-T.}\ \bibnamefont {Janka}},\ }\bibfield  {title}
  {\bibinfo {title} {{Fast neutrino flavor conversions in one-dimensional
  core-collapse supernova models with and without muon creation}},\ }\href
  {https://doi.org/10.1103/PhysRevD.103.063013} {\bibfield  {journal} {\bibinfo
   {journal} {Phys. Rev. D}\ }\textbf {\bibinfo {volume} {103}},\ \bibinfo
  {pages} {063013} (\bibinfo {year} {2021})},\ \Eprint
  {https://arxiv.org/abs/2012.08525} {arXiv:2012.08525 [astro-ph.HE]}
  \BibitemShut {NoStop}%
\bibitem [{\citenamefont {Dasgupta}\ \emph {et~al.}(2018)\citenamefont
  {Dasgupta}, \citenamefont {Mirizzi},\ and\ \citenamefont
  {Sen}}]{Dasgupta:2018ulw}%
  \BibitemOpen
  \bibfield  {author} {\bibinfo {author} {\bibfnamefont {B.}~\bibnamefont
  {Dasgupta}}, \bibinfo {author} {\bibfnamefont {A.}~\bibnamefont {Mirizzi}},\
  and\ \bibinfo {author} {\bibfnamefont {M.}~\bibnamefont {Sen}},\ }\bibfield
  {title} {\bibinfo {title} {{Simple method of diagnosing fast flavor
  conversions of supernova neutrinos}},\ }\href
  {https://doi.org/10.1103/PhysRevD.98.103001} {\bibfield  {journal} {\bibinfo
  {journal} {Phys. Rev. D}\ }\textbf {\bibinfo {volume} {98}},\ \bibinfo
  {pages} {103001} (\bibinfo {year} {2018})},\ \Eprint
  {https://arxiv.org/abs/1807.03322} {arXiv:1807.03322 [hep-ph]} \BibitemShut
  {NoStop}%
\bibitem [{\citenamefont {Glas}\ \emph {et~al.}(2020)\citenamefont {Glas},
  \citenamefont {Janka}, \citenamefont {Capozzi}, \citenamefont {Sen},
  \citenamefont {Dasgupta}, \citenamefont {Mirizzi},\ and\ \citenamefont
  {Sigl}}]{Glas:2019ijo}%
  \BibitemOpen
  \bibfield  {author} {\bibinfo {author} {\bibfnamefont {R.}~\bibnamefont
  {Glas}}, \bibinfo {author} {\bibfnamefont {H.-T.}\ \bibnamefont {Janka}},
  \bibinfo {author} {\bibfnamefont {F.}~\bibnamefont {Capozzi}}, \bibinfo
  {author} {\bibfnamefont {M.}~\bibnamefont {Sen}}, \bibinfo {author}
  {\bibfnamefont {B.}~\bibnamefont {Dasgupta}}, \bibinfo {author}
  {\bibfnamefont {A.}~\bibnamefont {Mirizzi}},\ and\ \bibinfo {author}
  {\bibfnamefont {G.}~\bibnamefont {Sigl}},\ }\bibfield  {title} {\bibinfo
  {title} {{Fast Neutrino Flavor Instability in the Neutron-star Convection
  Layer of Three-dimensional Supernova Models}},\ }\href
  {https://doi.org/10.1103/PhysRevD.101.063001} {\bibfield  {journal} {\bibinfo
   {journal} {Phys. Rev. D}\ }\textbf {\bibinfo {volume} {101}},\ \bibinfo
  {pages} {063001} (\bibinfo {year} {2020})},\ \Eprint
  {https://arxiv.org/abs/1912.00274} {arXiv:1912.00274 [astro-ph.HE]}
  \BibitemShut {NoStop}%
\bibitem [{\citenamefont {Nagakura}\ \emph {et~al.}(2021)\citenamefont
  {Nagakura}, \citenamefont {Johns}, \citenamefont {Burrows},\ and\
  \citenamefont {Fuller}}]{Nagakura:2021hyb}%
  \BibitemOpen
  \bibfield  {author} {\bibinfo {author} {\bibfnamefont {H.}~\bibnamefont
  {Nagakura}}, \bibinfo {author} {\bibfnamefont {L.}~\bibnamefont {Johns}},
  \bibinfo {author} {\bibfnamefont {A.}~\bibnamefont {Burrows}},\ and\ \bibinfo
  {author} {\bibfnamefont {G.~M.}\ \bibnamefont {Fuller}},\ }\bibfield  {title}
  {\bibinfo {title} {{Where, when, and why: Occurrence of fast-pairwise
  collective neutrino oscillation in three-dimensional core-collapse supernova
  models}},\ }\href {https://doi.org/10.1103/PhysRevD.104.083025} {\bibfield
  {journal} {\bibinfo  {journal} {Phys. Rev. D}\ }\textbf {\bibinfo {volume}
  {104}},\ \bibinfo {pages} {083025} (\bibinfo {year} {2021})},\ \Eprint
  {https://arxiv.org/abs/2108.07281} {arXiv:2108.07281 [astro-ph.HE]}
  \BibitemShut {NoStop}%
\bibitem [{\citenamefont {Johns}\ and\ \citenamefont
  {Nagakura}(2021)}]{Johns:2021taz}%
  \BibitemOpen
  \bibfield  {author} {\bibinfo {author} {\bibfnamefont {L.}~\bibnamefont
  {Johns}}\ and\ \bibinfo {author} {\bibfnamefont {H.}~\bibnamefont
  {Nagakura}},\ }\bibfield  {title} {\bibinfo {title} {{Fast flavor
  instabilities and the search for neutrino angular crossings}},\ }\href
  {https://doi.org/10.1103/PhysRevD.103.123012} {\bibfield  {journal} {\bibinfo
   {journal} {Phys. Rev. D}\ }\textbf {\bibinfo {volume} {103}},\ \bibinfo
  {pages} {123012} (\bibinfo {year} {2021})},\ \Eprint
  {https://arxiv.org/abs/2104.04106} {arXiv:2104.04106 [hep-ph]} \BibitemShut
  {NoStop}%
\bibitem [{\citenamefont {Cornelius}\ \emph {et~al.}(2025)\citenamefont
  {Cornelius}, \citenamefont {Tamborra}, \citenamefont {Heinlein},\ and\
  \citenamefont {Janka}}]{Cornelius:2025tyt}%
  \BibitemOpen
  \bibfield  {author} {\bibinfo {author} {\bibfnamefont {M.}~\bibnamefont
  {Cornelius}}, \bibinfo {author} {\bibfnamefont {I.}~\bibnamefont {Tamborra}},
  \bibinfo {author} {\bibfnamefont {M.}~\bibnamefont {Heinlein}},\ and\
  \bibinfo {author} {\bibfnamefont {H.-T.}\ \bibnamefont {Janka}},\ }\bibfield
  {title} {\bibinfo {title} {{Diagnosing electron-neutrino lepton number
  crossings in core-collapse supernovae: A comparison of methods}},\ }\href
  {https://doi.org/10.1103/gqd7-4ynz} {\bibfield  {journal} {\bibinfo
  {journal} {Phys. Rev. D}\ }\textbf {\bibinfo {volume} {112}},\ \bibinfo
  {pages} {063004} (\bibinfo {year} {2025})},\ \Eprint
  {https://arxiv.org/abs/2506.20723} {arXiv:2506.20723 [astro-ph.HE]}
  \BibitemShut {NoStop}%
\bibitem [{\citenamefont {Shalgar}\ and\ \citenamefont
  {Tamborra}(2024{\natexlab{a}})}]{Shalgar:2023aca}%
  \BibitemOpen
  \bibfield  {author} {\bibinfo {author} {\bibfnamefont {S.}~\bibnamefont
  {Shalgar}}\ and\ \bibinfo {author} {\bibfnamefont {I.}~\bibnamefont
  {Tamborra}},\ }\bibfield  {title} {\bibinfo {title} {{Do neutrinos become
  flavor unstable due to collisions with matter in the supernova decoupling
  region?}},\ }\href {https://doi.org/10.1103/PhysRevD.109.103011} {\bibfield
  {journal} {\bibinfo  {journal} {Phys. Rev. D}\ }\textbf {\bibinfo {volume}
  {109}},\ \bibinfo {pages} {103011} (\bibinfo {year} {2024}{\natexlab{a}})},\
  \Eprint {https://arxiv.org/abs/2307.10366} {arXiv:2307.10366 [astro-ph.HE]}
  \BibitemShut {NoStop}%
\bibitem [{\citenamefont {Shalgar}\ and\ \citenamefont
  {Tamborra}(2024{\natexlab{b}})}]{Shalgar:2024gjt}%
  \BibitemOpen
  \bibfield  {author} {\bibinfo {author} {\bibfnamefont {S.}~\bibnamefont
  {Shalgar}}\ and\ \bibinfo {author} {\bibfnamefont {I.}~\bibnamefont
  {Tamborra}},\ }\bibfield  {title} {\bibinfo {title} {{Neutrino quantum
  kinetics in a core-collapse supernova}},\ }\href
  {https://doi.org/10.1088/1475-7516/2024/09/021} {\bibfield  {journal}
  {\bibinfo  {journal} {JCAP}\ }\textbf {\bibinfo {volume} {09}},\ \bibinfo
  {pages} {021}},\ \Eprint {https://arxiv.org/abs/2406.09504} {arXiv:2406.09504
  [astro-ph.HE]} \BibitemShut {NoStop}%
\bibitem [{\citenamefont {Shalgar}\ and\ \citenamefont
  {Tamborra}(2025)}]{Shalgar:2025oht}%
  \BibitemOpen
  \bibfield  {author} {\bibinfo {author} {\bibfnamefont {S.}~\bibnamefont
  {Shalgar}}\ and\ \bibinfo {author} {\bibfnamefont {I.}~\bibnamefont
  {Tamborra}},\ }\href@noop {} {\bibinfo {title} {{Neutrino quantum kinetics in
  three flavors}}} (\bibinfo {year} {2025}),\ \Eprint
  {https://arxiv.org/abs/2503.03835} {arXiv:2503.03835 [astro-ph.HE]}
  \BibitemShut {NoStop}%
\bibitem [{\citenamefont {{The Garching Core-Collapse Supernova
  Archive}}()}]{Garching_CCSN_archive}%
  \BibitemOpen
  \bibfield  {author} {\bibinfo {author} {\bibnamefont {{The Garching
  Core-Collapse Supernova Archive}}},\ }\href@noop {} {}\bibinfo {howpublished}
  {\url{https://wwwmpa.mpa-garching.mpg.de/ccsnarchive/archive.html}}\BibitemShut
  {NoStop}%
\bibitem [{\citenamefont {Lattimer}\ and\ \citenamefont
  {Swesty}(1991)}]{Lattimer:1991nc}%
  \BibitemOpen
  \bibfield  {author} {\bibinfo {author} {\bibfnamefont {J.~M.}\ \bibnamefont
  {Lattimer}}\ and\ \bibinfo {author} {\bibfnamefont {F.~D.}\ \bibnamefont
  {Swesty}},\ }\bibfield  {title} {\bibinfo {title} {{A Generalized equation of
  state for hot, dense matter}},\ }\href
  {https://doi.org/10.1016/0375-9474(91)90452-C} {\bibfield  {journal}
  {\bibinfo  {journal} {Nucl. Phys. A}\ }\textbf {\bibinfo {volume} {535}},\
  \bibinfo {pages} {331} (\bibinfo {year} {1991})}\BibitemShut {NoStop}%
\bibitem [{\citenamefont {Steiner}\ \emph {et~al.}(2013)\citenamefont
  {Steiner}, \citenamefont {Hempel},\ and\ \citenamefont
  {Fischer}}]{Steiner:2012rk}%
  \BibitemOpen
  \bibfield  {author} {\bibinfo {author} {\bibfnamefont {A.~W.}\ \bibnamefont
  {Steiner}}, \bibinfo {author} {\bibfnamefont {M.}~\bibnamefont {Hempel}},\
  and\ \bibinfo {author} {\bibfnamefont {T.}~\bibnamefont {Fischer}},\
  }\bibfield  {title} {\bibinfo {title} {{Core-collapse supernova equations of
  state based on neutron star observations}},\ }\href
  {https://doi.org/10.1088/0004-637X/774/1/17} {\bibfield  {journal} {\bibinfo
  {journal} {Astrophys. J.}\ }\textbf {\bibinfo {volume} {774}},\ \bibinfo
  {pages} {17} (\bibinfo {year} {2013})},\ \Eprint
  {https://arxiv.org/abs/1207.2184} {arXiv:1207.2184 [astro-ph.SR]}
  \BibitemShut {NoStop}%
\bibitem [{\citenamefont {Hempel}\ and\ \citenamefont
  {Schaffner-Bielich}(2010)}]{Hempel:2009mc}%
  \BibitemOpen
  \bibfield  {author} {\bibinfo {author} {\bibfnamefont {M.}~\bibnamefont
  {Hempel}}\ and\ \bibinfo {author} {\bibfnamefont {J.}~\bibnamefont
  {Schaffner-Bielich}},\ }\bibfield  {title} {\bibinfo {title} {{Statistical
  Model for a Complete Supernova Equation of State}},\ }\href
  {https://doi.org/10.1016/j.nuclphysa.2010.02.010} {\bibfield  {journal}
  {\bibinfo  {journal} {Nucl. Phys. A}\ }\textbf {\bibinfo {volume} {837}},\
  \bibinfo {pages} {210} (\bibinfo {year} {2010})},\ \Eprint
  {https://arxiv.org/abs/0911.4073} {arXiv:0911.4073 [nucl-th]} \BibitemShut
  {NoStop}%
\bibitem [{\citenamefont {Typel}\ \emph {et~al.}(2010)\citenamefont {Typel},
  \citenamefont {R{\"o}pke}, \citenamefont {Klahn}, \citenamefont {Blaschke},\
  and\ \citenamefont {Wolter}}]{Typel:2009sy}%
  \BibitemOpen
  \bibfield  {author} {\bibinfo {author} {\bibfnamefont {S.}~\bibnamefont
  {Typel}}, \bibinfo {author} {\bibfnamefont {G.}~\bibnamefont {R{\"o}pke}},
  \bibinfo {author} {\bibfnamefont {T.}~\bibnamefont {Klahn}}, \bibinfo
  {author} {\bibfnamefont {D.}~\bibnamefont {Blaschke}},\ and\ \bibinfo
  {author} {\bibfnamefont {H.~H.}\ \bibnamefont {Wolter}},\ }\bibfield  {title}
  {\bibinfo {title} {{Composition and thermodynamics of nuclear matter with
  light clusters}},\ }\href {https://doi.org/10.1103/PhysRevC.81.015803}
  {\bibfield  {journal} {\bibinfo  {journal} {Phys. Rev. C}\ }\textbf {\bibinfo
  {volume} {81}},\ \bibinfo {pages} {015803} (\bibinfo {year} {2010})},\
  \Eprint {https://arxiv.org/abs/0908.2344} {arXiv:0908.2344 [nucl-th]}
  \BibitemShut {NoStop}%
\bibitem [{\citenamefont {Hempel}\ \emph {et~al.}(2012)\citenamefont {Hempel},
  \citenamefont {Fischer}, \citenamefont {Schaffner-Bielich},\ and\
  \citenamefont {Liebend{\"o}rfer}}]{Hempel:2011mk}%
  \BibitemOpen
  \bibfield  {author} {\bibinfo {author} {\bibfnamefont {M.}~\bibnamefont
  {Hempel}}, \bibinfo {author} {\bibfnamefont {T.}~\bibnamefont {Fischer}},
  \bibinfo {author} {\bibfnamefont {J.}~\bibnamefont {Schaffner-Bielich}},\
  and\ \bibinfo {author} {\bibfnamefont {M.}~\bibnamefont {Liebend{\"o}rfer}},\
  }\bibfield  {title} {\bibinfo {title} {{New Equations of State in Simulations
  of Core-Collapse Supernovae}},\ }\href
  {https://doi.org/10.1088/0004-637X/748/1/70} {\bibfield  {journal} {\bibinfo
  {journal} {Astrophys. J.}\ }\textbf {\bibinfo {volume} {748}},\ \bibinfo
  {pages} {70} (\bibinfo {year} {2012})},\ \Eprint
  {https://arxiv.org/abs/1108.0848} {arXiv:1108.0848 [astro-ph.HE]}
  \BibitemShut {NoStop}%
\bibitem [{\citenamefont {Bollig}\ \emph {et~al.}(2017)\citenamefont {Bollig},
  \citenamefont {Janka}, \citenamefont {Lohs}, \citenamefont
  {Mart{\'i}nez-Pinedo}, \citenamefont {Horowitz},\ and\ \citenamefont
  {Melson}}]{Bollig:2017lki}%
  \BibitemOpen
  \bibfield  {author} {\bibinfo {author} {\bibfnamefont {R.}~\bibnamefont
  {Bollig}}, \bibinfo {author} {\bibfnamefont {H.-T.}\ \bibnamefont {Janka}},
  \bibinfo {author} {\bibfnamefont {A.}~\bibnamefont {Lohs}}, \bibinfo {author}
  {\bibfnamefont {G.}~\bibnamefont {Mart{\'i}nez-Pinedo}}, \bibinfo {author}
  {\bibfnamefont {C.~J.}\ \bibnamefont {Horowitz}},\ and\ \bibinfo {author}
  {\bibfnamefont {T.}~\bibnamefont {Melson}},\ }\bibfield  {title} {\bibinfo
  {title} {{Muon Creation in Supernova Matter Facilitates Neutrino-driven
  Explosions}},\ }\href {https://doi.org/10.1103/PhysRevLett.119.242702}
  {\bibfield  {journal} {\bibinfo  {journal} {Phys. Rev. Lett.}\ }\textbf
  {\bibinfo {volume} {119}},\ \bibinfo {pages} {242702} (\bibinfo {year}
  {2017})},\ \Eprint {https://arxiv.org/abs/1706.04630} {arXiv:1706.04630
  [astro-ph.HE]} \BibitemShut {NoStop}%
\bibitem [{\citenamefont {{Roberts}}\ \emph {et~al.}(2012)\citenamefont
  {{Roberts}}, \citenamefont {{Shen}}, \citenamefont {{Cirigliano}},
  \citenamefont {{Pons}}, \citenamefont {{Reddy}},\ and\ \citenamefont
  {{Woosley}}}]{2012PhRvL.108f1103R}%
  \BibitemOpen
  \bibfield  {author} {\bibinfo {author} {\bibfnamefont {L.~F.}\ \bibnamefont
  {{Roberts}}}, \bibinfo {author} {\bibfnamefont {G.}~\bibnamefont {{Shen}}},
  \bibinfo {author} {\bibfnamefont {V.}~\bibnamefont {{Cirigliano}}}, \bibinfo
  {author} {\bibfnamefont {J.~A.}\ \bibnamefont {{Pons}}}, \bibinfo {author}
  {\bibfnamefont {S.}~\bibnamefont {{Reddy}}},\ and\ \bibinfo {author}
  {\bibfnamefont {S.~E.}\ \bibnamefont {{Woosley}}},\ }\bibfield  {title}
  {\bibinfo {title} {{Protoneutron Star Cooling with Convection: The Effect of
  the Symmetry Energy}},\ }\href
  {https://doi.org/10.1103/PhysRevLett.108.061103} {\bibfield  {journal}
  {\bibinfo  {journal} {Phys. Rev. Lett.}\ }\textbf {\bibinfo {volume} {108}},\
  \bibinfo {eid} {061103} (\bibinfo {year} {2012})},\ \Eprint
  {https://arxiv.org/abs/1112.0335} {arXiv:1112.0335 [astro-ph.HE]}
  \BibitemShut {NoStop}%
\bibitem [{\citenamefont {Pascal}\ \emph {et~al.}(2022)\citenamefont {Pascal},
  \citenamefont {Novak},\ and\ \citenamefont {Oertel}}]{Pascal:2022qeg}%
  \BibitemOpen
  \bibfield  {author} {\bibinfo {author} {\bibfnamefont {A.}~\bibnamefont
  {Pascal}}, \bibinfo {author} {\bibfnamefont {J.}~\bibnamefont {Novak}},\ and\
  \bibinfo {author} {\bibfnamefont {M.}~\bibnamefont {Oertel}},\ }\bibfield
  {title} {\bibinfo {title} {{Proto-neutron star evolution with improved
  charged-current neutrino\textendash{}nucleon interactions}},\ }\href
  {https://doi.org/10.1093/mnras/stac016} {\bibfield  {journal} {\bibinfo
  {journal} {Mon. Not. Roy. Astron. Soc.}\ }\textbf {\bibinfo {volume} {511}},\
  \bibinfo {pages} {356} (\bibinfo {year} {2022})},\ \Eprint
  {https://arxiv.org/abs/2201.01955} {arXiv:2201.01955 [nucl-th]} \BibitemShut
  {NoStop}%
\bibitem [{\citenamefont {Mirizzi}\ \emph {et~al.}(2016)\citenamefont
  {Mirizzi}, \citenamefont {Tamborra}, \citenamefont {Janka}, \citenamefont
  {Saviano}, \citenamefont {Scholberg}, \citenamefont {Bollig}, \citenamefont
  {H{\"u}depohl},\ and\ \citenamefont {Chakraborty}}]{Mirizzi:2015eza}%
  \BibitemOpen
  \bibfield  {author} {\bibinfo {author} {\bibfnamefont {A.}~\bibnamefont
  {Mirizzi}}, \bibinfo {author} {\bibfnamefont {I.}~\bibnamefont {Tamborra}},
  \bibinfo {author} {\bibfnamefont {H.-T.}\ \bibnamefont {Janka}}, \bibinfo
  {author} {\bibfnamefont {N.}~\bibnamefont {Saviano}}, \bibinfo {author}
  {\bibfnamefont {K.}~\bibnamefont {Scholberg}}, \bibinfo {author}
  {\bibfnamefont {R.}~\bibnamefont {Bollig}}, \bibinfo {author} {\bibfnamefont
  {L.}~\bibnamefont {H{\"u}depohl}},\ and\ \bibinfo {author} {\bibfnamefont
  {S.}~\bibnamefont {Chakraborty}},\ }\bibfield  {title} {\bibinfo {title}
  {{Supernova Neutrinos: Production, Oscillations and Detection}},\ }\href
  {https://doi.org/10.1393/ncr/i2016-10120-8} {\bibfield  {journal} {\bibinfo
  {journal} {Riv. Nuovo Cim.}\ }\textbf {\bibinfo {volume} {39}},\ \bibinfo
  {pages} {1} (\bibinfo {year} {2016})},\ \Eprint
  {https://arxiv.org/abs/1508.00785} {arXiv:1508.00785 [astro-ph.HE]}
  \BibitemShut {NoStop}%
\bibitem [{\citenamefont {Rampp}\ and\ \citenamefont
  {Janka}(2002)}]{Rampp:2002bq}%
  \BibitemOpen
  \bibfield  {author} {\bibinfo {author} {\bibfnamefont {M.}~\bibnamefont
  {Rampp}}\ and\ \bibinfo {author} {\bibfnamefont {H.-T.}\ \bibnamefont
  {Janka}},\ }\bibfield  {title} {\bibinfo {title} {{Radiation hydrodynamics
  with neutrinos: Variable Eddington factor method for core collapse supernova
  simulations}},\ }\href {https://doi.org/10.1051/0004-6361:20021398}
  {\bibfield  {journal} {\bibinfo  {journal} {Astron. Astrophys.}\ }\textbf
  {\bibinfo {volume} {396}},\ \bibinfo {pages} {361} (\bibinfo {year}
  {2002})},\ \Eprint {https://arxiv.org/abs/astro-ph/0203101}
  {arXiv:astro-ph/0203101} \BibitemShut {NoStop}%
\bibitem [{\citenamefont {Fiorillo}\ \emph {et~al.}(2023)\citenamefont
  {Fiorillo}, \citenamefont {Heinlein}, \citenamefont {Janka}, \citenamefont
  {Raffelt}, \citenamefont {Vitagliano},\ and\ \citenamefont
  {Bollig}}]{Fiorillo:2023frv}%
  \BibitemOpen
  \bibfield  {author} {\bibinfo {author} {\bibfnamefont {D.~F.~G.}\
  \bibnamefont {Fiorillo}}, \bibinfo {author} {\bibfnamefont {M.}~\bibnamefont
  {Heinlein}}, \bibinfo {author} {\bibfnamefont {H.-T.}\ \bibnamefont {Janka}},
  \bibinfo {author} {\bibfnamefont {G.~G.}\ \bibnamefont {Raffelt}}, \bibinfo
  {author} {\bibfnamefont {E.}~\bibnamefont {Vitagliano}},\ and\ \bibinfo
  {author} {\bibfnamefont {R.}~\bibnamefont {Bollig}},\ }\bibfield  {title}
  {\bibinfo {title} {{Supernova simulations confront SN 1987A neutrinos}},\
  }\href {https://doi.org/10.1103/PhysRevD.108.083040} {\bibfield  {journal}
  {\bibinfo  {journal} {Phys. Rev. D}\ }\textbf {\bibinfo {volume} {108}},\
  \bibinfo {pages} {083040} (\bibinfo {year} {2023})},\ \Eprint
  {https://arxiv.org/abs/2308.01403} {arXiv:2308.01403 [astro-ph.HE]}
  \BibitemShut {NoStop}%
\bibitem [{\citenamefont {Sigl}\ and\ \citenamefont
  {Raffelt}(1993)}]{Sigl:1993ctk}%
  \BibitemOpen
  \bibfield  {author} {\bibinfo {author} {\bibfnamefont {G.}~\bibnamefont
  {Sigl}}\ and\ \bibinfo {author} {\bibfnamefont {G.~G.}\ \bibnamefont
  {Raffelt}},\ }\bibfield  {title} {\bibinfo {title} {{General kinetic
  description of relativistic mixed neutrinos}},\ }\href
  {https://doi.org/10.1016/0550-3213(93)90175-O} {\bibfield  {journal}
  {\bibinfo  {journal} {Nucl. Phys. B}\ }\textbf {\bibinfo {volume} {406}},\
  \bibinfo {pages} {423} (\bibinfo {year} {1993})}\BibitemShut {NoStop}%
\bibitem [{\citenamefont {O'Connor}(2015)}]{OConnor:2014sgn}%
  \BibitemOpen
  \bibfield  {author} {\bibinfo {author} {\bibfnamefont {E.}~\bibnamefont
  {O'Connor}},\ }\bibfield  {title} {\bibinfo {title} {{An Open-Source Neutrino
  Radiation Hydrodynamics Code for Core-Collapse Supernovae}},\ }\href
  {https://doi.org/10.1088/0067-0049/219/2/24} {\bibfield  {journal} {\bibinfo
  {journal} {Astrophys. J. Suppl.}\ }\textbf {\bibinfo {volume} {219}},\
  \bibinfo {pages} {24} (\bibinfo {year} {2015})},\ \Eprint
  {https://arxiv.org/abs/1411.7058} {arXiv:1411.7058 [astro-ph.HE]}
  \BibitemShut {NoStop}%
\bibitem [{\citenamefont {Buras}\ \emph {et~al.}(2006)\citenamefont {Buras},
  \citenamefont {Rampp}, \citenamefont {Janka},\ and\ \citenamefont
  {Kifonidis}}]{Buras:2005rp}%
  \BibitemOpen
  \bibfield  {author} {\bibinfo {author} {\bibfnamefont {R.}~\bibnamefont
  {Buras}}, \bibinfo {author} {\bibfnamefont {M.}~\bibnamefont {Rampp}},
  \bibinfo {author} {\bibfnamefont {H.-T.}\ \bibnamefont {Janka}},\ and\
  \bibinfo {author} {\bibfnamefont {K.}~\bibnamefont {Kifonidis}},\ }\bibfield
  {title} {\bibinfo {title} {{Two-dimensional hydrodynamic core-collapse
  supernova simulations with spectral neutrino transport. 1. Numerical method
  and results for a 15 solar mass star}},\ }\href
  {https://doi.org/10.1051/0004-6361:20053783} {\bibfield  {journal} {\bibinfo
  {journal} {Astron. Astrophys.}\ }\textbf {\bibinfo {volume} {447}},\ \bibinfo
  {pages} {1049} (\bibinfo {year} {2006})},\ \Eprint
  {https://arxiv.org/abs/astro-ph/0507135} {arXiv:astro-ph/0507135}
  \BibitemShut {NoStop}%
\bibitem [{\citenamefont {Janka}\ \emph {et~al.}(2007)\citenamefont {Janka},
  \citenamefont {Langanke}, \citenamefont {Marek}, \citenamefont
  {Mart{\'i}nez-Pinedo},\ and\ \citenamefont {M{\"u}ller}}]{Janka:2006fh}%
  \BibitemOpen
  \bibfield  {author} {\bibinfo {author} {\bibfnamefont {H.-T.}\ \bibnamefont
  {Janka}}, \bibinfo {author} {\bibfnamefont {K.}~\bibnamefont {Langanke}},
  \bibinfo {author} {\bibfnamefont {A.}~\bibnamefont {Marek}}, \bibinfo
  {author} {\bibfnamefont {G.}~\bibnamefont {Mart{\'i}nez-Pinedo}},\ and\
  \bibinfo {author} {\bibfnamefont {B.}~\bibnamefont {M{\"u}ller}},\ }\bibfield
   {title} {\bibinfo {title} {{Theory of Core-Collapse Supernovae}},\ }\href
  {https://doi.org/10.1016/j.physrep.2007.02.002} {\bibfield  {journal}
  {\bibinfo  {journal} {Phys. Rept.}\ }\textbf {\bibinfo {volume} {442}},\
  \bibinfo {pages} {38} (\bibinfo {year} {2007})},\ \Eprint
  {https://arxiv.org/abs/astro-ph/0612072} {arXiv:astro-ph/0612072}
  \BibitemShut {NoStop}%
\bibitem [{\citenamefont {Janka}(2012)}]{Janka:2012wk}%
  \BibitemOpen
  \bibfield  {author} {\bibinfo {author} {\bibfnamefont {H.-T.}\ \bibnamefont
  {Janka}},\ }\bibfield  {title} {\bibinfo {title} {{Explosion Mechanisms of
  Core-Collapse Supernovae}},\ }\href
  {https://doi.org/10.1146/annurev-nucl-102711-094901} {\bibfield  {journal}
  {\bibinfo  {journal} {Ann. Rev. Nucl. Part. Sci.}\ }\textbf {\bibinfo
  {volume} {62}},\ \bibinfo {pages} {407} (\bibinfo {year} {2012})},\ \Eprint
  {https://arxiv.org/abs/1206.2503} {arXiv:1206.2503 [astro-ph.SR]}
  \BibitemShut {NoStop}%
\bibitem [{\citenamefont {Delfan~Azari}\ \emph {et~al.}(2020)\citenamefont
  {Delfan~Azari}, \citenamefont {Yamada}, \citenamefont {Morinaga},
  \citenamefont {Nagakura}, \citenamefont {Furusawa}, \citenamefont {Harada},
  \citenamefont {Okawa}, \citenamefont {Iwakami},\ and\ \citenamefont
  {Sumiyoshi}}]{DelfanAzari:2019tez}%
  \BibitemOpen
  \bibfield  {author} {\bibinfo {author} {\bibfnamefont {M.}~\bibnamefont
  {Delfan~Azari}}, \bibinfo {author} {\bibfnamefont {S.}~\bibnamefont
  {Yamada}}, \bibinfo {author} {\bibfnamefont {T.}~\bibnamefont {Morinaga}},
  \bibinfo {author} {\bibfnamefont {H.}~\bibnamefont {Nagakura}}, \bibinfo
  {author} {\bibfnamefont {S.}~\bibnamefont {Furusawa}}, \bibinfo {author}
  {\bibfnamefont {A.}~\bibnamefont {Harada}}, \bibinfo {author} {\bibfnamefont
  {H.}~\bibnamefont {Okawa}}, \bibinfo {author} {\bibfnamefont
  {W.}~\bibnamefont {Iwakami}},\ and\ \bibinfo {author} {\bibfnamefont
  {K.}~\bibnamefont {Sumiyoshi}},\ }\bibfield  {title} {\bibinfo {title} {{Fast
  collective neutrino oscillations inside the neutrino sphere in core-collapse
  supernovae}},\ }\href {https://doi.org/10.1103/PhysRevD.101.023018}
  {\bibfield  {journal} {\bibinfo  {journal} {Phys. Rev. D}\ }\textbf {\bibinfo
  {volume} {101}},\ \bibinfo {pages} {023018} (\bibinfo {year} {2020})},\
  \Eprint {https://arxiv.org/abs/1910.06176} {arXiv:1910.06176 [astro-ph.HE]}
  \BibitemShut {NoStop}%
\bibitem [{\citenamefont {Akaho}\ \emph {et~al.}(2023)\citenamefont {Akaho},
  \citenamefont {Harada}, \citenamefont {Nagakura}, \citenamefont {Iwakami},
  \citenamefont {Okawa}, \citenamefont {Furusawa}, \citenamefont {Matsufuru},
  \citenamefont {Sumiyoshi},\ and\ \citenamefont {Yamada}}]{Akaho:2022zdz}%
  \BibitemOpen
  \bibfield  {author} {\bibinfo {author} {\bibfnamefont {R.}~\bibnamefont
  {Akaho}}, \bibinfo {author} {\bibfnamefont {A.}~\bibnamefont {Harada}},
  \bibinfo {author} {\bibfnamefont {H.}~\bibnamefont {Nagakura}}, \bibinfo
  {author} {\bibfnamefont {W.}~\bibnamefont {Iwakami}}, \bibinfo {author}
  {\bibfnamefont {H.}~\bibnamefont {Okawa}}, \bibinfo {author} {\bibfnamefont
  {S.}~\bibnamefont {Furusawa}}, \bibinfo {author} {\bibfnamefont
  {H.}~\bibnamefont {Matsufuru}}, \bibinfo {author} {\bibfnamefont
  {K.}~\bibnamefont {Sumiyoshi}},\ and\ \bibinfo {author} {\bibfnamefont
  {S.}~\bibnamefont {Yamada}},\ }\bibfield  {title} {\bibinfo {title}
  {{Protoneutron Star Convection Simulated with a New General Relativistic
  Boltzmann Neutrino Radiation Hydrodynamics Code}},\ }\href
  {https://doi.org/10.3847/1538-4357/acad76} {\bibfield  {journal} {\bibinfo
  {journal} {Astrophys. J.}\ }\textbf {\bibinfo {volume} {944}},\ \bibinfo
  {pages} {60} (\bibinfo {year} {2023})},\ \Eprint
  {https://arxiv.org/abs/2206.01673} {arXiv:2206.01673 [astro-ph.HE]}
  \BibitemShut {NoStop}%
\bibitem [{\citenamefont {Akaho}\ \emph {et~al.}(2024)\citenamefont {Akaho},
  \citenamefont {Liu}, \citenamefont {Nagakura}, \citenamefont {Zaizen},\ and\
  \citenamefont {Yamada}}]{Akaho:2023brj}%
  \BibitemOpen
  \bibfield  {author} {\bibinfo {author} {\bibfnamefont {R.}~\bibnamefont
  {Akaho}}, \bibinfo {author} {\bibfnamefont {J.}~\bibnamefont {Liu}}, \bibinfo
  {author} {\bibfnamefont {H.}~\bibnamefont {Nagakura}}, \bibinfo {author}
  {\bibfnamefont {M.}~\bibnamefont {Zaizen}},\ and\ \bibinfo {author}
  {\bibfnamefont {S.}~\bibnamefont {Yamada}},\ }\bibfield  {title} {\bibinfo
  {title} {{Collisional and fast neutrino flavor instabilities in
  two-dimensional core-collapse supernova simulation with Boltzmann neutrino
  transport}},\ }\href {https://doi.org/10.1103/PhysRevD.109.023012} {\bibfield
   {journal} {\bibinfo  {journal} {Phys. Rev. D}\ }\textbf {\bibinfo {volume}
  {109}},\ \bibinfo {pages} {023012} (\bibinfo {year} {2024})},\ \Eprint
  {https://arxiv.org/abs/2311.11272} {arXiv:2311.11272 [astro-ph.HE]}
  \BibitemShut {NoStop}%
\bibitem [{\citenamefont {Nagakura}\ \emph {et~al.}(2019)\citenamefont
  {Nagakura}, \citenamefont {Morinaga}, \citenamefont {Kato},\ and\
  \citenamefont {Yamada}}]{Nagakura:2019sig}%
  \BibitemOpen
  \bibfield  {author} {\bibinfo {author} {\bibfnamefont {H.}~\bibnamefont
  {Nagakura}}, \bibinfo {author} {\bibfnamefont {T.}~\bibnamefont {Morinaga}},
  \bibinfo {author} {\bibfnamefont {C.}~\bibnamefont {Kato}},\ and\ \bibinfo
  {author} {\bibfnamefont {S.}~\bibnamefont {Yamada}},\ }\bibfield  {title}
  {\bibinfo {title} {{Fast-pairwise collective neutrino oscillations associated
  with asymmetric neutrino emissions in core-collapse supernova}},\ }\href
  {https://doi.org/10.3847/1538-4357/ab4cf2} {\bibfield  {journal} {\bibinfo
  {journal} {The Astrophysical Journal}\ }\textbf {\bibinfo {volume} {886}},\
  \bibinfo {pages} {139} (\bibinfo {year} {2019})},\ \Eprint
  {https://arxiv.org/abs/1910.04288} {arXiv:1910.04288 [astro-ph.HE]}
  \BibitemShut {NoStop}%
\bibitem [{\citenamefont {Morinaga}\ \emph {et~al.}(2020)\citenamefont
  {Morinaga}, \citenamefont {Nagakura}, \citenamefont {Kato},\ and\
  \citenamefont {Yamada}}]{Morinaga:2019wsv}%
  \BibitemOpen
  \bibfield  {author} {\bibinfo {author} {\bibfnamefont {T.}~\bibnamefont
  {Morinaga}}, \bibinfo {author} {\bibfnamefont {H.}~\bibnamefont {Nagakura}},
  \bibinfo {author} {\bibfnamefont {C.}~\bibnamefont {Kato}},\ and\ \bibinfo
  {author} {\bibfnamefont {S.}~\bibnamefont {Yamada}},\ }\bibfield  {title}
  {\bibinfo {title} {{Fast neutrino-flavor conversion in the preshock region of
  core-collapse supernovae}},\ }\href
  {https://doi.org/10.1103/PhysRevResearch.2.012046} {\bibfield  {journal}
  {\bibinfo  {journal} {Phys. Rev. Res.}\ }\textbf {\bibinfo {volume} {2}},\
  \bibinfo {pages} {012046} (\bibinfo {year} {2020})},\ \Eprint
  {https://arxiv.org/abs/1909.13131} {arXiv:1909.13131 [astro-ph.HE]}
  \BibitemShut {NoStop}%
\bibitem [{\citenamefont {Harada}\ and\ \citenamefont
  {Nagakura}(2022)}]{Harada:2021ata}%
  \BibitemOpen
  \bibfield  {author} {\bibinfo {author} {\bibfnamefont {A.}~\bibnamefont
  {Harada}}\ and\ \bibinfo {author} {\bibfnamefont {H.}~\bibnamefont
  {Nagakura}},\ }\bibfield  {title} {\bibinfo {title} {{Prospects of Fast
  Flavor Neutrino Conversion in Rotating Core-collapse Supernovae}},\ }\href
  {https://doi.org/10.3847/1538-4357/ac38a0} {\bibfield  {journal} {\bibinfo
  {journal} {Astrophys. J.}\ }\textbf {\bibinfo {volume} {924}},\ \bibinfo
  {pages} {109} (\bibinfo {year} {2022})},\ \Eprint
  {https://arxiv.org/abs/2110.08291} {arXiv:2110.08291 [astro-ph.HE]}
  \BibitemShut {NoStop}%
\bibitem [{\citenamefont {Tamborra}\ \emph {et~al.}(2014)\citenamefont
  {Tamborra}, \citenamefont {Hanke}, \citenamefont {Janka}, \citenamefont
  {M\"uller}, \citenamefont {Raffelt},\ and\ \citenamefont
  {Marek}}]{Tamborra:2014aua}%
  \BibitemOpen
  \bibfield  {author} {\bibinfo {author} {\bibfnamefont {I.}~\bibnamefont
  {Tamborra}}, \bibinfo {author} {\bibfnamefont {F.}~\bibnamefont {Hanke}},
  \bibinfo {author} {\bibfnamefont {H.-T.}\ \bibnamefont {Janka}}, \bibinfo
  {author} {\bibfnamefont {B.}~\bibnamefont {M\"uller}}, \bibinfo {author}
  {\bibfnamefont {G.~G.}\ \bibnamefont {Raffelt}},\ and\ \bibinfo {author}
  {\bibfnamefont {A.}~\bibnamefont {Marek}},\ }\bibfield  {title} {\bibinfo
  {title} {{Self-sustained asymmetry of lepton-number emission: A new
  phenomenon during the supernova shock-accretion phase in three dimensions}},\
  }\href {https://doi.org/10.1088/0004-637X/792/2/96} {\bibfield  {journal}
  {\bibinfo  {journal} {Astrophys. J.}\ }\textbf {\bibinfo {volume} {792}},\
  \bibinfo {pages} {96} (\bibinfo {year} {2014})},\ \Eprint
  {https://arxiv.org/abs/1402.5418} {arXiv:1402.5418 [astro-ph.SR]}
  \BibitemShut {NoStop}%
\bibitem [{\citenamefont {Wu}\ and\ \citenamefont
  {Tamborra}(2017)}]{Wu:2017qpc}%
  \BibitemOpen
  \bibfield  {author} {\bibinfo {author} {\bibfnamefont {M.-R.}\ \bibnamefont
  {Wu}}\ and\ \bibinfo {author} {\bibfnamefont {I.}~\bibnamefont {Tamborra}},\
  }\bibfield  {title} {\bibinfo {title} {{Fast neutrino conversions: Ubiquitous
  in compact binary merger remnants}},\ }\href
  {https://doi.org/10.1103/PhysRevD.95.103007} {\bibfield  {journal} {\bibinfo
  {journal} {Phys. Rev. D}\ }\textbf {\bibinfo {volume} {95}},\ \bibinfo
  {pages} {103007} (\bibinfo {year} {2017})},\ \Eprint
  {https://arxiv.org/abs/1701.06580} {arXiv:1701.06580 [astro-ph.HE]}
  \BibitemShut {NoStop}%
\bibitem [{\citenamefont {Lund}\ \emph {et~al.}(2025)\citenamefont {Lund},
  \citenamefont {Mukhopadhyay}, \citenamefont {Miller},\ and\ \citenamefont
  {McLaughlin}}]{Lund:2025jjo}%
  \BibitemOpen
  \bibfield  {author} {\bibinfo {author} {\bibfnamefont {K.~A.}\ \bibnamefont
  {Lund}}, \bibinfo {author} {\bibfnamefont {P.}~\bibnamefont {Mukhopadhyay}},
  \bibinfo {author} {\bibfnamefont {J.~M.}\ \bibnamefont {Miller}},\ and\
  \bibinfo {author} {\bibfnamefont {G.~C.}\ \bibnamefont {McLaughlin}},\
  }\bibfield  {title} {\bibinfo {title} {{Angle-dependent in Situ Fast Flavor
  Transformations in Post-neutron-star-merger Disks}},\ }\href
  {https://doi.org/10.3847/2041-8213/add0a7} {\bibfield  {journal} {\bibinfo
  {journal} {Astrophys. J. Lett.}\ }\textbf {\bibinfo {volume} {985}},\
  \bibinfo {pages} {L9} (\bibinfo {year} {2025})},\ \Eprint
  {https://arxiv.org/abs/2503.23727} {arXiv:2503.23727 [astro-ph.HE]}
  \BibitemShut {NoStop}%
\bibitem [{\citenamefont {Mukhopadhyay}\ \emph {et~al.}(2024)\citenamefont
  {Mukhopadhyay}, \citenamefont {Miller},\ and\ \citenamefont
  {McLaughlin}}]{Mukhopadhyay:2024zzl}%
  \BibitemOpen
  \bibfield  {author} {\bibinfo {author} {\bibfnamefont {P.}~\bibnamefont
  {Mukhopadhyay}}, \bibinfo {author} {\bibfnamefont {J.}~\bibnamefont
  {Miller}},\ and\ \bibinfo {author} {\bibfnamefont {G.~C.}\ \bibnamefont
  {McLaughlin}},\ }\bibfield  {title} {\bibinfo {title} {{The Time Evolution of
  Fast Flavor Crossings in Postmerger Disks around a Black Hole Remnant}},\
  }\href {https://doi.org/10.3847/1538-4357/ad6c42} {\bibfield  {journal}
  {\bibinfo  {journal} {Astrophys. J.}\ }\textbf {\bibinfo {volume} {974}},\
  \bibinfo {pages} {110} (\bibinfo {year} {2024})},\ \Eprint
  {https://arxiv.org/abs/2404.17938} {arXiv:2404.17938 [astro-ph.HE]}
  \BibitemShut {NoStop}%
\end{thebibliography}%

\end{document}